\documentclass[reprint,superscriptaddress,amsmath,amssymb,aps,pre]{revtex4-2}
\usepackage{graphicx}
\usepackage{dcolumn}
\usepackage{bm}
\usepackage{color}
\usepackage{units}
\usepackage{wasysym}
\usepackage{MnSymbol}
\usepackage{comment}
\usepackage{times}
\usepackage[unicode=true,pdfusetitle,bookmarks=false,colorlinks=true,citecolor=blue,urlcolor=blue,linkcolor=blue]{hyperref}
\begin{document}

\title{Phase Coexistence Implications of Violating Newton's Third Law}

\author{Yu-Jen Chiu}
\affiliation{Department of Materials Science and Engineering, University of California, Berkeley, California 94720, USA}

\author{Ahmad K. Omar}
\email{aomar@berkeley.edu}
\affiliation{Department of Materials Science and Engineering, University of California, Berkeley, California 94720, USA}
\affiliation{Materials Sciences Division, Lawrence Berkeley National Laboratory, Berkeley, California 94720, USA}

\begin{abstract}
Newton's third law, \textit{action = reaction}, is a foundational statement of classical mechanics. 
However, in natural and living systems, this law appears to be routinely violated for constituents interacting in a nonequilibrium environment. 
Here, we use computer simulations to explore the macroscopic phase behavior implications of breaking microscopic interaction reciprocity for a simple model system. 
We consider a binary mixture of attractive particles and introduce a parameter that is a continuous measure of the degree to which interaction reciprocity is broken. 
In the reciprocal limit, the species are indistinguishable and the system phase separates into domains with distinct densities and identical compositions.
Increasing nonreciprocity is found to drive the system to explore a rich assortment of phases, including phases with strong composition asymmetries and three-phase coexistence. 
Many of the states induced by these forces, including traveling crystals and liquids, have no equilibrium analogue. 
By mapping the complete phase diagram for this model system and characterizing these unique phases, our findings offer a concrete path forward towards understanding how nonreciprocity shapes the structures found in living systems and how this might be leveraged in the design of synthetic materials.
\end{abstract}

\maketitle

\textit{Introduction.--} 
Living and natural systems across all length scales can appear to violate Newton's third law. 
A familiar example is the so-called ``predator-prey'' interaction, whereby one entity (the predator) feels an attractive force toward the other, while the other (the prey) is repelled.
These \textit{effective nonreciprocal interactions} can emerge from a host of complex factors and can have far-reaching implications on collective phenomena, phase transitions, and pattern formation.
In the living world, examples range from the interactions amongst bacteria~\cite{Long2001, Czirok1996} at the microscale to the dynamics of animal herds~\cite{Vicsek2012,Strandburg2013} at the macroscale.
Synthetic systems with nonreciprocal interactions have also been increasingly realized, with nonreciprocity emerging in systems as diverse as dusty plasmas~\cite{Tsytovich1997,Chaudhuri2011,Morfill2009,Keh2016, Ivlev2017,Kompaneets2016}, colloidal suspensions~\cite{Meredith2020,Sriram2012,Soto2014, Saha2019}, and even solid metamaterials~\cite{Scheibner2020, Miri2019, Nassar2020}. 
The understanding of the structure and phases of biological systems and leveraging nonreciprocal interactions in synthetic materials will require a fundamental understanding of both the origins and implications of these effective interactions. 

Effective forces acting on particles of interest have long been known to shape the structure and properties of condensed matter systems.
These forces emerge from the coarse-graining of degrees of freedom, such as other species present within the system.
When the coarse-grained degrees of freedom are in equilibrium the structure of the effective forces is severely restricted: the one-body forces (e.g.,~fluctuating Brownian and dissipative forces acting on the solute upon coarse-graining a molecular solvent) must satisfy the fluctuation-dissipation theorem (FDT) and many-body forces (e.g.,~the pairwise depletion force between larger solutes upon coarse-graining smaller solutes) must be conservative and consistent with Newton's third law~\cite{Israelachvili1992, Dijkstra2000, Bolhuis2001, Praprotnik2008, Mognetti2009}.  

Coarse-graining degrees of freedom that are out of equilibrium unlocks a new range of possibilities. 
In a nonequilibrium environment, one-body forces need not satisfy FDT with these ``active'' forces, resulting in novel phase transitions and collective phenomena~\cite{Cates2015, Bechinger2016}.    
A driven environment may also generate effective \textit{nonreciprocal} many-body forces~\cite{Hayashi2006,Durve2018,Ivlev2015,Agudo2019,You2020,Saha2020,Fruchart2021,Saha2022,Osat2022,Dinelli2022}. 
The consequences of violating interaction reciprocity on phase behavior remain unclear (especially in comparison to those of one-body active forces) and have motivated the development of a variety of theoretical perspectives~\cite{Ivlev2015,You2020,Agudo2019,Saha2020,Saha2022,Fruchart2021,Dinelli2022}. 

In this article, we explore the consequences of violating Newton's third law on the phase behavior of a model system where nonreciprocity can be continuously tuned, defining an additional axis on the phase diagram.
Increasing nonreciprocity not only leads to exotic states of coexistence, but may also \textit{stabilize} homogeneous states resembling active fluids.
Our findings thus reveal that nonreciprocal interactions may impact the phase behavior in new and unexpected ways and constitute an important step forward toward the development of a multicomponent nonequilibrium coexistence theory. 

\begin{figure*}
	\centering
	\includegraphics[width=.95\textwidth]{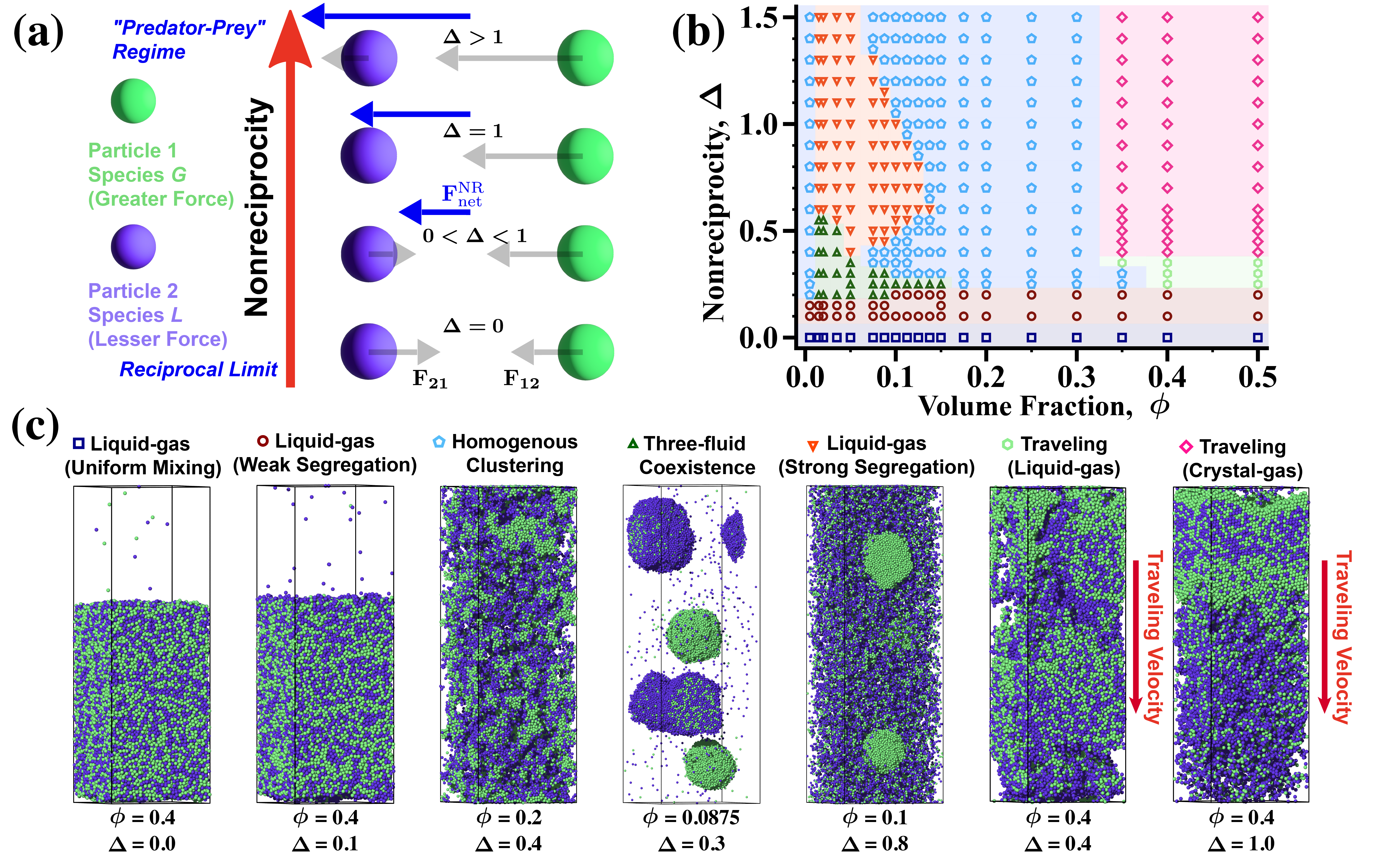}
	\caption{\protect\small{{(a) Schematic of pairwise nonreciprocal interactions between species $L$ and $G$. (b) Phase diagram in the density-nonreciprocity plane with $\varepsilon/k_{B}T = 2.0$ and $\chi = 0.5$, with $\phi \in [0.005, 0.5], \Delta \in [0, 1.5]$. (c) Representative snapshots of each of the seven distinct regions of the phase diagram.
 }}}
	\label{fig:phasediagram}
\end{figure*}

\textit{Model system.--} 
Elucidating the influence of nonreciprocal interactions on material phase behavior requires a model system in which the degree of nonreciprocity can be continuously varied.
We consider a binary mixture of particles of species $L$ and $G$ with the pairwise force exerted on particle $i$ by particle $j$ taking the following form~\cite{Ivlev2015, Bartnick2015, Ivlev2017, Kryuchkov2018}:
\begin{equation}
\label{eq:modelforces}
\mathbf{F}_{ij}(\mathbf{r}) = 
\mathbf{F}_{ij}^{\rm C}(\mathbf{r}) \times
\begin{cases}
[1-\Delta(r)] \; & \text{, $ij$ $\in$ $LG$} \\
[1+\Delta(r)] \; & \text{, $ij$ $\in$ $GL$}\\
1 & \text{, $ij$ $\in$ $LL$ or $GG$}
\end{cases}
\end{equation}
where $\mathbf{r}$ is the interparticle distance vector (with magnitude $r$), $\mathbf{F}_{ij}^{\rm C}(\mathbf{r})$ is a \textit{conservative} interaction force satisfying Newton's third law $\left({\rm i.e.,}~\mathbf{F}_{ij}^{\rm C} + \mathbf{F}_{ji}^{\rm C} = \mathbf{0}\right)$, and $\Delta(r)$ is a nonreciprocity function.
Interactions between particles of like species are entirely reciprocal $\left(\mathbf{F}_{ij} = \mathbf{F}_{ij}^{\rm C}\right)$ while interspecies interactions exhibit both reciprocal $\mathbf{F}_{ij}^{\rm R} = \mathbf{F}^{\rm C}_{ij}$ and nonreciprocal $\mathbf{F}_{ij}^{\rm NR} = \pm\Delta \times \mathbf{F}^{\rm C}_{ij}$ contributions with $\mathbf{F}_{ij} = \mathbf{F}_{ij}^{\rm R} + \mathbf{F}_{ij}^{\rm NR}$. 
Consider $\text{$ij$ $\in$ $LG$}$: The sum of this pair interaction will result in a net force (violating Newton's third law) with $\mathbf{F}_{ij} + \mathbf{F}_{ji} = 2\mathbf{F}_{ij}^{\rm NR} = 2\Delta\mathbf{F}^{\rm C}_{ij}$.
This resulting net force $\mathbf{F}^{\rm NR}_{\rm net} = 2\mathbf{F}^{\rm NR}_{ij}$ [see Fig.~\ref{fig:phasediagram}(a)] may be thought of as an internally generated active force that entirely depends on particle configurations and the nonreciprocity function $\Delta(r)$. 
For simplicity, we consider a step nonreciprocity function $\Delta(r) = \Delta \Theta(r-d_{\rm rec})$ such that interactions between particles are entirely reciprocal for separation distances less than a reciprocity diameter, $d_{\rm rec}$, and have a constant degree of nonreciprocity $\Delta$ for distances greater than $d_{\rm rec}$.
{\color{black}This model now allows us to continuously depart from the equilibrium limit ($\Delta = 0$) and isolate the precise role of the violation of interaction reciprocity on a systems' phase diagram. 
In this sense, this model represents a minimal system in which detailed balance is broken on a \textit{two-body} level, joining a family of simple models in which detailed-balance is broken with a single parameter~\cite{Bechinger2016, Omar2022, Kumar2020, Agrawal2022}.
}
 
Figure~\ref{fig:phasediagram}(a) illustrates the qualitatively distinct interaction regimes controlled by $\Delta$.
Departures from $\Delta = 0$ break interaction reciprocity, resulting in forces that are no longer conservative and a nonequilibrium distribution of microstates. 
For $0<\Delta<1$, while the magnitude of the force is different for each species, the forces continue to oppose each other: repulsive (attractive) interactions will continue to be repulsive (attractive). 
However, for $\Delta > 1$, the nonreciprocity results in forces that no longer oppose each other: a particle exerting a repulsive (attractive) force will itself experience an attractive (repulsive) force [see Fig.~\ref{fig:phasediagram}(a)].
This is the so-called ``predator-prey" interaction that appears in nature.
While this might lead one to conclude that $\Delta=1$ is a significant dynamical point from the individual particle perspective, from the standpoint of the \textit{particle pair}, this point is not unique. 
The net nonreciprocal force between the particle pair [see Fig.~\ref{fig:phasediagram}(a)] linearly grows with increasing $\Delta$: there is nothing to distinguish $\Delta = 1$.

\begin{figure*}
	\centering
	\includegraphics[width=.95\textwidth]{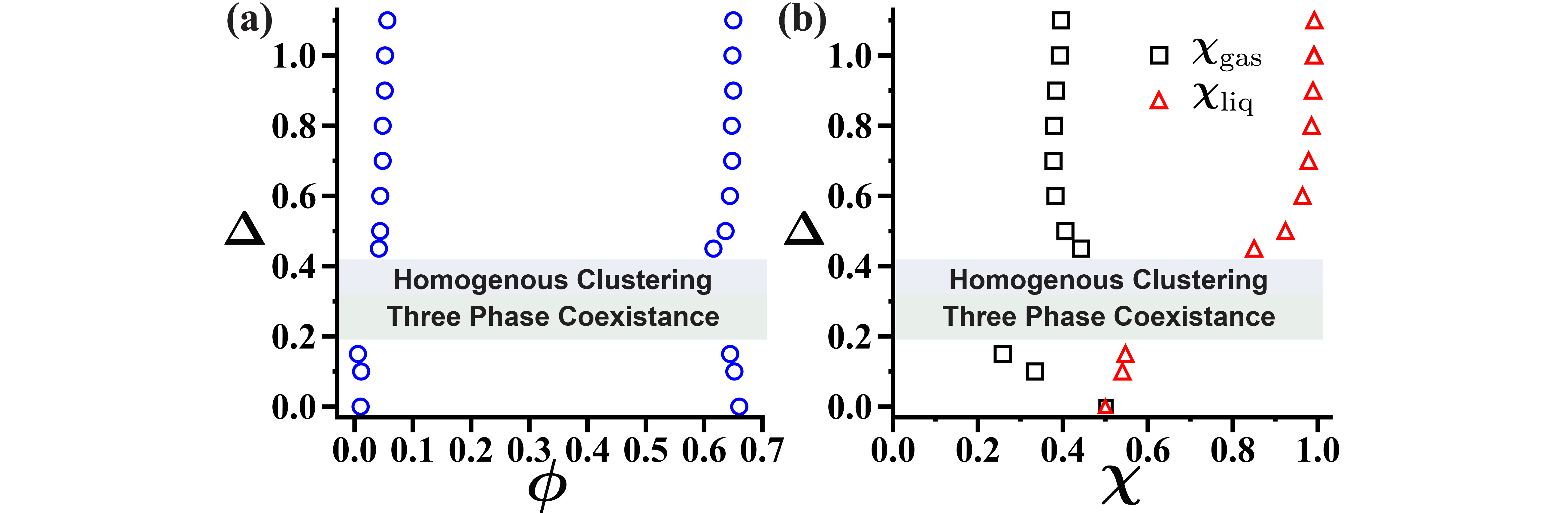}
	\caption{\protect\small{{Nonreciprocity dependence of coexisting (a) densities and (b) compositions for states of two-phase coexistence with $\phi = 0.0875$.}}}
	\label{fig:twophase}
\end{figure*}

Our aim is to identify how nonreciprocity shapes the structural and dynamical landscape of a simple model system. 
To this end, we take our conservative force $\mathbf{F}^{C}$ to result from Lennard-Jones (LJ) interactions (with a cutoff distance of $2.5\sigma$), introducing the LJ energy ($\varepsilon$) and length ($\sigma$) scales.
The reciprocal diameter is set to $d_{\rm rec} = 2^{1/6}\sigma$ such that all particle pairs experience strictly reciprocal repulsion within separation distances of $d_{\rm rec}$.
{\color{black}The resulting pair interaction forces are shown in Fig.~\ref{fig:forces} in the Appendix.}
The particle dynamics follow the overdamped Langevin equation (see the Appendix), which imparts an ideal translational diffusivity $D_T = k_BT/\zeta$ (introducing our system timescale $\tau = \sigma^2/D_T$) where $k_BT$ is the thermal energy and $\zeta$ is the translational drag coefficient.
Our system state is thus fully described by four parameters: the degree of nonreciprocity $\Delta$, the global volume fraction $\phi = \left(\rho_L + \rho_G\right)\pi(d_{\rm rec})^3/6$ (where $\rho_L$ and $\rho_G$ are the number densities of species $L$ and $G$, respectively), the global composition $\chi = \rho_G/(\rho_L+\rho_G)$, and the ratio of the interaction energy to the thermal energy $\varepsilon/k_{B}T$.
In this work, we fix $\chi = 0.5$ and $\varepsilon/k_{B}T = 2.0$, and systematically sweep $\phi$ and $\Delta$.
All simulations were conducted with $50000$ particles using \texttt{HOOMD-blue}~\cite{Anderson2020}.

\textit{Phase diagram.--} 
Figure~\ref{fig:phasediagram}(b) displays the phase diagram obtained from extensive computer simulations.
{\color{black}For finite nonreciprocity, the system is out of equilibrium, and, while there has been recent progress in the sampling of driven systems~\cite{Klymko2018, Ray2018, Das2019a, Whitelam2020, Ray2020, Helms2020, Oakes2020, Rose2021, Das2022},} we address questions of global stability by conducting a number of long-time simulations with distinct initial configurations, as detailed in the Supporting Information (SI)~\footnote{See Supporting Information, which includes Refs.~\cite{Martinez2009, Stukowski2009, Weber2016, Hargus2021, Weeks1971}, for additional simulation and calculation details as well as simulation movies.}.

\textit{Stationary phase transitions.--} 
In the reciprocal limit $\Delta = 0$, there is no distinction between $L$ and $G$ particles. 
The stationary state corresponds to that of a single-component attractive Lennard-Jones system at $\varepsilon/k_BT = 2.0$.
While it has been established~\cite{Schultz2018} that the thermodynamic ground state for this system is crystal (fcc)-fluid coexistence, observing this state requires an exceedingly rare fluctuation, to generate the critical nuclei necessary for crystal growth.
A long-lived (metastable) liquid-gas coexistence is, instead, observed for all densities examined in this work with $\Delta = 0$. 

Small departures from the reciprocal limit ($0<\Delta\leq 0.15$) continue to result in liquid-gas coexistence with similar (slightly reduced) coexisting densities [see Fig.~\ref{fig:twophase}(a)]. 
However, a composition asymmetry between the phases is now generated as shown in Fig.~\ref{fig:twophase}(b).
For small $\Delta$, it may be permissible to assume that the system remains in a local equilibrium with modified interaction potentials between particle pairs. 
However, a composition distinction between phases in a symmetric mixture would require difference in the modified $LL$ and $GG$ interaction energies. 
As nonreciprocity leaves interactions between like species unaltered, systems with weak nonreciprocal forces cannot be mapped to equilibrium with modified interaction energies.

An alternative intriguing possibility exists for this weak species segregation in which, for small $\Delta$, rather than mapping the system to an equilibrium system with modified interactions, we consider the effective temperatures of each species to slightly deviate from the bath temperature and with $\left(T_L \neq T_G \right)$.
The statistical mechanics of systems with particles in contact with different heat baths is considerably simpler than a full treatment of nonreciprocity.
In fact, this perspective was found to hold exactly in the limit of small nonreciprocity for certain model systems~\cite{Ivlev2015}. However, its general applicability to nonreciprocal systems remains to be determined.

\begin{figure*}
	\centering
	\includegraphics[width=.95\textwidth]{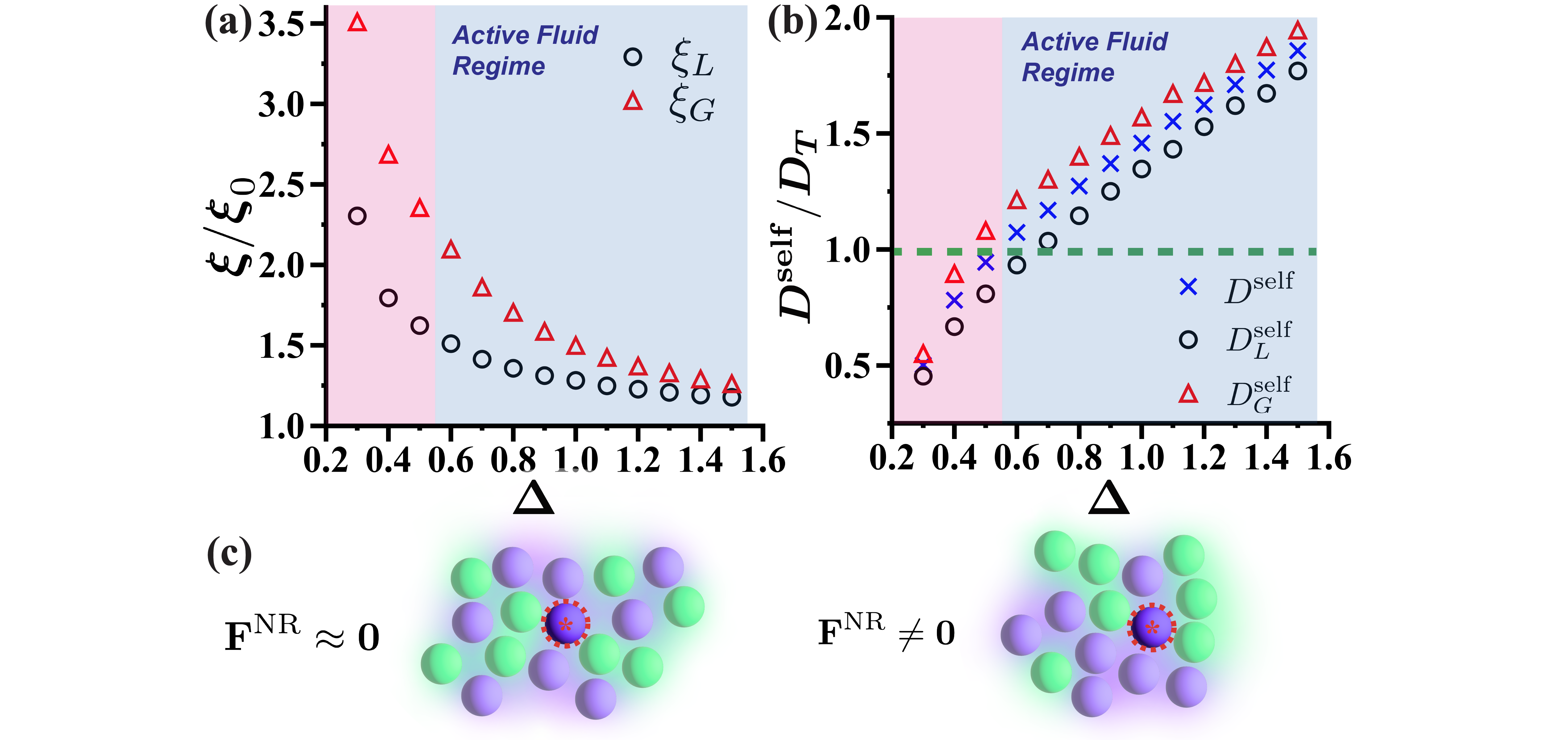}
	\caption{\protect\small{{Nonreciprocity dependence of the (a) correlation lengths [normalized by the ideal gas value $\xi_0$ (see the Appendix)] and (b) self diffusivities for $\phi=0.2$ and $\Delta \geq 0.3$. Shaded regions highlight the active fluid regime ($D^{\rm self} > D_T$). (c) Fluctuations of the local composition environment of a tagged particle control nonreciprocal forcing.}}}
	\label{fig:clustering}
\end{figure*}

In scenarios with species of different temperatures, particles with a higher effective temperature are entropically driven to enrich the dilute phase~\cite{Weber2016,Note1}. 
Our simulation results (Fig.~\ref{fig:twophase}) show that particles of species $L$ are biased toward the gaseous phases, suggesting $T_L>T_G$ if this perspective holds.
However, there is little evidence to suggest that $T_L > T_G$ as the self diffusion of $G$ is always greater than that of $L$ [see Fig.~\ref{fig:clustering}(b)]. 
Attempting to define the effective temperature with a Stokes-Einstein relation (i.e.,~$k_BT_{L/G} = D^{\rm self}_{L/G}\zeta^{\rm eff}$, where $\zeta^{\rm eff}$ is an effective translational drag coefficient, assumed to be identical for each species) would lead to the interpretation that $T_G > T_L (> T)$.
We, thus, conclude that even for the small values of nonreciprocity examined here, effective equilibrium and effective temperature ideas are not sufficient to describe the observed phase behavior.

Increasing the nonreciprocity beyond $\Delta > 0.15$ results in a multitude of states, depending on the overall volume fraction. 
For a broad range of intermediate densities $\left(0.15 < \phi < 0.35\right)$, the system is globally homogeneous (i.e., there is no phase separation). 
This broad region of intermediate densities at which no phase transitions occur $\left (0.15 < \phi < 0.35\right )$ is not the only region of homogeneity. 
For $\Delta \ge 0.2$, we indeed find the system to be homogeneous at the lowest concentration ($\phi = 0.005$) shown in the phase diagram presented in Fig.~\ref{fig:phasediagram}(b), while systems with $\Delta < 0.2$ are phase separated.
It is interesting that even at these low concentrations, the system state remains $\Delta$ dependent.
In the low density limit as $\phi \rightarrow 0$ this $\Delta$ dependence must vanish, as interactions become negligible and the system is indistinguishable from an ideal gas with temperature $k_BT$. 

In this broad region of homogeneity at intermediate densities, the system consists of clusters of finite spatial extent.
Each cluster is typically enriched in either $L$ or $G$, and as a result, the system appears to be microphase separated, particularly at smaller values of $\Delta$ [cf.,~Fig.~\ref{fig:phasediagram}(c)].
We characterize the extent of species segregation by computing the partial static structure factors, provided in the SI~\cite{Note1}. 
The correlation lengths for density fluctuations of species $L$ and $G$, $\xi_{L}$ and $\xi_{G}$, respectively, are extracted from the structure factor and shown in Fig.~\ref{fig:clustering}(a). 
With increasing $\Delta$, we find a monotonic reduction in these correlation lengths: increasing nonreciprocity homogenizes the system at these densities. 
We note that the divergence of both $\xi_{L}$ and $\xi_{G}$ as $\Delta \rightarrow 0.2$ is anticipated as the phase-separated regime is approached in this limit. 

What is more, the dramatic alteration to the system's structural properties is accompanied by a commensurate change in the species' self diffusivity [see Fig.~\ref{fig:clustering}(b)].
The diffusion constant in the absence of any interactions is the ideal Stokes-Einstein translational diffusivity, $D_T = k_BT/\zeta$, imparted on the particles by the Langevin bath. 
In the case of passive systems with purely reciprocal interactions, $D_T$ serves as an upper bound for the self-diffusion constant with $D^{\rm self} \le D_T$. 
Furthermore, if a Stokes-Einstein relation holds, $D^{\rm self} = k_BT/\zeta^{\rm eff}$ where $\zeta^{\rm eff} \geq \zeta$.
Upon increasing $\Delta$, the self diffusivities of both species increase significantly and even \textit{exceed the ideal diffusivity} for $\Delta \gtrsim 0.6$.

We can determine the precise origins of this enhanced diffusion using the Green-Kubo relation for self-diffusion~\cite{Hargus2021} and leveraging the substitution of forces for velocities permitted by overdamped dynamics.
Composition fluctuations about a tagged particle [see Fig.~\ref{fig:clustering}(c)] generate instantaneous nonreciprocal forces that act to increase the diffusivity (similar to a one-body active force), and the system resembles an active fluid. 
However, this contribution is found to be relatively independent of $\Delta$, as detailed in the SI~\cite{Note1}.
The origin of the dramatic increase in particle diffusivity with $\Delta$ is in fact found to be the result of the diminishing magnitude of a number of force correlations that reduce particle mobility~\cite{Note1}. 
Increased nonreciprocal forcing induces profound structural changes that have a clear dynamical consequence: the breakup of segregated microphases reduces the effective particle friction.  
The diminished friction with increasing $\Delta$ coupled with the direct contribution of nonreciprocal forcing to particle diffusion, is what allows the system to exceed the ideal diffusion. 

\begin{figure*}
	\centering
	\includegraphics[width=.95\textwidth]{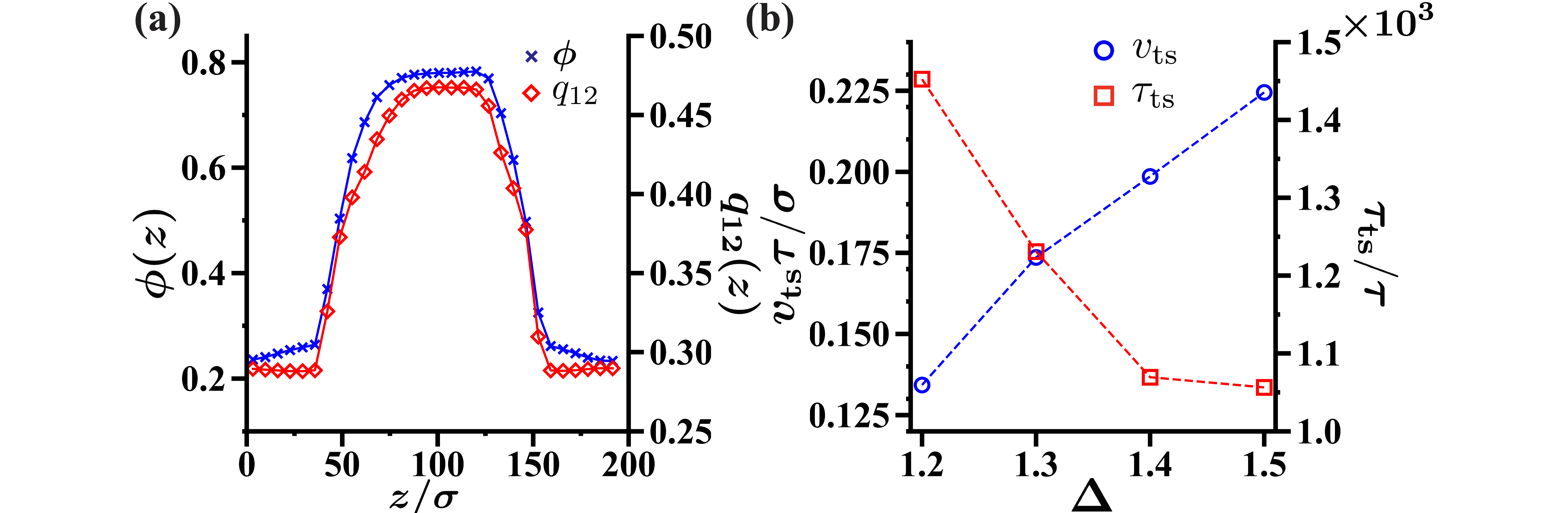}
	\caption{\protect\small{{Traveling crystal (a) long time-averaged $\phi$ and $q_{\rm 12}$ profiles (averaging time $\gg \tau_{\rm ts}$) with $\left(\Delta=1.5, \phi=0.5\right)$ and (b) $\Delta$ dependence of the traveling state dynamics for $\phi = 0.5$.}}} 
	\label{fig:traveling}
\end{figure*}

At volume fractions between regions of global homogeneity [i.e.,~$0.005 < \phi \leq 0.15$], nonreciprocity introduces a variety of possible stationary states.
Appreciable values of nonreciprocity $\left(0.2 \leq \Delta \leq 0.55\right)$ results in a state of \textit{three-phase coexistence}, as shown in Fig.~\ref{fig:phasediagram}(b).
The three coexisting fluids consist of two high density spherical liquid phases~\footnote{The low overall density results in these dense liquids occupying little volume, forming spherical droplets within the dilute fluid phase which comprises most of the system by volume.}, each nearly pure in species $L$ or $G$, and a dilute phase with a relatively equal amount of each species.
In contrast to nonreciprocal two-phase coexistence, the observed densities and compositions may more readily be mapped to an equilibrium analog.    
In equilibrium, reducing the magnitude of interspecies attraction, i.e.,~$\varepsilon_{LG} < \varepsilon$, can generate a state of three-phase coexistence not dissimilar to what we observe here, as detailed in the SI~\cite{Note1}.

Further increasing $\Delta$, will result in a return to two-fluid coexistence, with densities similar to those seen at low nonreciprocity [see Fig.~\ref{fig:twophase}(a)].
However, unlike lower values of nonreciprocity, there is a sharp composition contrast between the coexisting fluids, as shown in Fig.~\ref{fig:twophase}(b). 
Intriguingly, the dense fluid [again adopting a spherical morphology, see Fig.~\ref{fig:phasediagram}(c)] is nearly entirely comprised of species $G$. 
Preparing a system seeded with a liquid droplet, purely of species $L$, in this two-phase region of the phase diagram again results in the nucleation and growth of a liquid $G$ droplet -- dense liquids of $G$ appear to be the preferred state~\cite{Note1}.
In single-component, nonequilibrium systems, interfacial mechanics uniquely determine the coexisting densities~\cite{Aifantis1983a, Omar2022}. 
While this remains to be generalized to multicomponent systems, the interfacial nonreciprocal forces result in a pure $G$ liquid phase coexisting with a dilute mixture as the mechanically stable state of coexistence.   

With increasing nonreciprocity, this region of two-phase coexistence, with strong species segregation, is found to narrow, while the stable homogeneous region of the phase diagram expands [see Fig.~\ref{fig:phasediagram}(b)]. 
At the highest value of nonreciprocity reported here, the phase diagram has four distinct regions. 
Beginning at the lowest density, with increasing $\phi$, we move from the homogeneous ``ideal gas" regime to a two-fluid coexistence, followed by a broad regime of homogeneity. 
Further increasing the density beyond $\phi \geq 0.35$ results in a traveling state-a dynamical transition that we now discuss. 

\textit{Traveling states.--} At high $\phi$, performing the long-time average results in density and composition profiles that resemble traditional states of two-phase coexistence. 
Intriguingly, at these high concentrations and with increasing nonreciprocity, the dense phase undergoes an ordering transition, with the emergence of an fcc crystal, quantified with the Steinhardt-Nelson-Ronchetti order parameter, measuring 12-fold rotational symmetry, $q_{12}$ (see the Appendix)~\cite{Steinhardt1983}.
Typical density and $q_{12}$ profiles are shown in Fig.~\ref{fig:traveling}(a), with $z$ being the direction normal to the interface.
Did the mobility imparted by nonreciprocal forces allow the system to surmount the nucleation barrier or is nonreciprocity itself driving this ordering transition?
We leave interrogating these questions for a future study, but recent work on active order-disorder transitions~\cite{Omar2021} suggests that nonequilibrium forcing can entirely reshape the crystallization landscape.

While the long-time averages of these states appear consistent with a stationary two-phase coexistence scenario, careful examination of the low-density region reveals a kink in its density profile, suggesting that this is not a typical state of coexistence between two homogeneous phases.
Indeed, observing the system dynamics reveals the emergence of a traveling state, defined by ballistic center-of-mass motion [see Fig.~\ref{fig:phasediagram}(c) and SI~\cite{Note1}].
This density gradient generates an asymmetry that drives a persistent center-of-mass velocity: the coexisting domains appear to chase each other.

These \textit{traveling states} occur for both liquid-gas and crystal-fluid coexistence, depending on the degree of nonreciprocity [see Fig.~\ref{fig:phasediagram}(b)].
Over a time period $\tau_{\rm ts}$, the direction of the low-density phases' density gradient reverses, and with it so too does the center-of-mass velocity $v_{\rm com}$.
These dynamics are well-described by a simple oscillatory form ${v_{\rm com}(t) = v_{\rm ts}\sin(2\pi t/\tau_{\rm ts})}$.
Figure~\ref{fig:traveling}(b) reveals that while the velocity increases nearly linearly with nonreciprocity, there is a marked decrease in the period~\footnote{We note that our simulation duration ($5000\tau $) limited the periods we could observe to those with $\tau _{\protect \rm ts} < 2500\tau $, which coincides with $\Delta \geq 1.2$.}.
That the velocity scales nearly linearly with $\Delta$ is perhaps indicative that the traveling speed simply scales with the magnitude of the nonreciprocal forcing while the oscillation period is dictated by the timescale for the reorganization of the density gradient. 
The enhanced mobility with increasing $\Delta$ [cf.,~Fig.~\ref{fig:clustering}(b)] likely drives the marked reduction in the gradient reorganization time, generating the observed rapid oscillation frequency at high nonreciprocity.
 
The density gradient, which drives the traveling state is also accompanied by a composition gradient: the velocity direction is normal to the interface enriched in $G$ [see Fig.~\ref{fig:phasediagram}(c) and SI movies].
These traveling states thus appear to be a version of the ``predator-prey'' interaction with the $G$ enriched regions chasing the $G$ depleted domains.
It is important to note that traveling states do not require nonreciprocal interactions. 
Mixtures of active and passive particles~\cite{Stenhammar2015, Wysocki2016, Wittkowski2017, Omar2019}, interacting with purely reciprocal interactions, can also exhibit such states~\cite{Wysocki2016, Wittkowski2017, You2020,Dinelli2022}. 

What determines the onset of these dynamical states is the mutual diffusion coefficient tensor $D_{\alpha\beta}$ the elements of which describe how a diffusive flux of species $\alpha$ emerges from a density gradient in species $\beta$~\footnote{The diagonal elements of the mutual diffusion tensor are not to be confused with the self diffusion coefficients computed in this work. 
The mutual diffusion tensor describes the system response to imposed gradients, and its elements are only equal to self-diffusion coefficients in the dilute limit.}.
Linear stability analysis using the species mass conservation equations results in stability criteria purely in terms of the eigenvalues of $D_{\alpha\beta}$. 
The traveling phases may emerge when the eigenvalues have imaginary components~\cite{Wittkowski2017,You2020,Saha2020,Saha2022,Dinelli2022}.
Constructing the stability diagram for this system will thus require a model for the diffusion tensor.
The mutual diffusion coefficients have been analytically computed for some active-passive mixtures~\cite{Wittkowski2017}, beginning from the evolution equation for the distribution of microscopic configurations for those systems.
A general microscopic expression for the mutual diffusion tensor for mixtures with nonreciprocal interactions is the subject of ongoing work, as it will be necessary for the prediction of traveling states from first principles.
While $D_{\alpha\beta}$ is crucial for determining stability criteria, predicting the properties of the stable states (e.g.,~the density and composition of phases) will require additional considerations.

\begin{figure*}
	\centering
	\includegraphics[width=.675\textwidth]{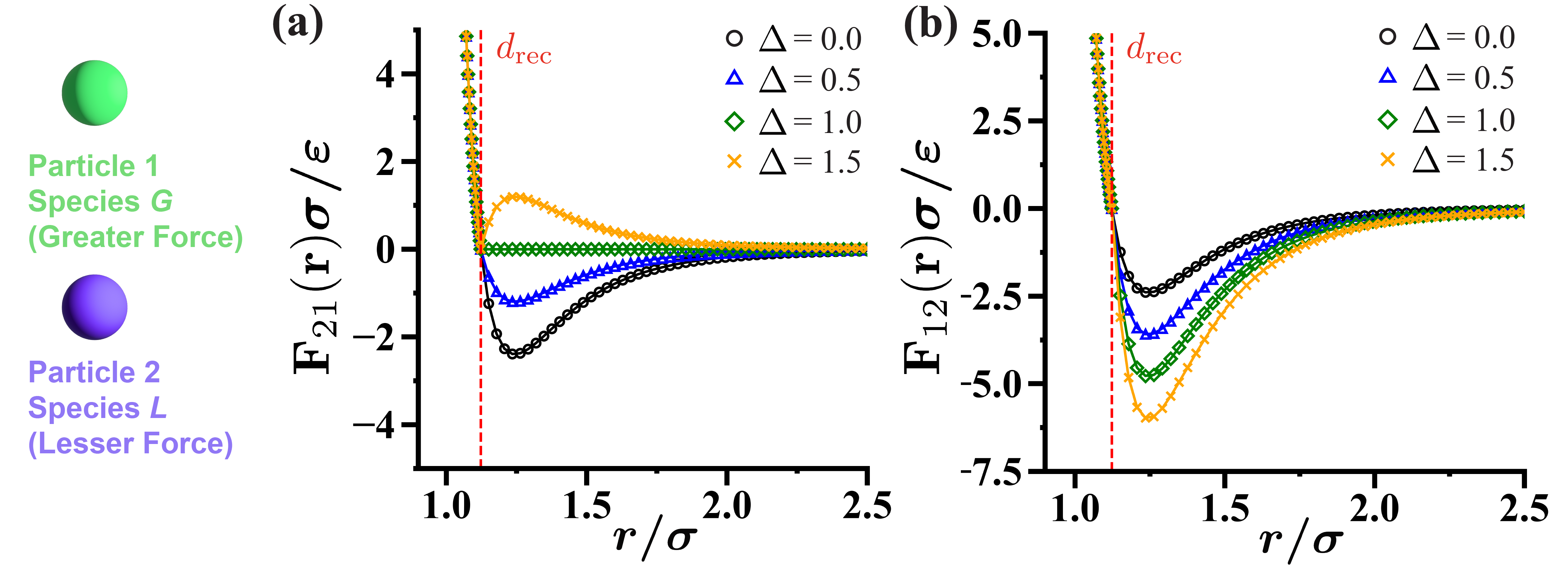}
	\caption{\protect\small{{$\Delta$ dependence of the pair interaction force acting on (a) a particle of species $L$ by a particle of species $G$ and (b) a particle of species $G$ by a particle of species $L$.}}}
	\label{fig:forces}
\end{figure*}

\textit{Conclusions.--} 
Nonreciprocity is a generic feature of living and natural systems and is increasingly used as a means to alter the structure and dynamics of synthetic materials across a multitude of length scales.
Recent works have sought to generalize statistical mechanics and the theory of dynamical systems to account for nonreciprocity~\cite{Ivlev2015,You2020,Agudo2019,Saha2020,Saha2022,Fruchart2021,Dinelli2022}, however, the theory for multicomponent nonequilibrium coexistence remains an outstanding challenge.
Here, we have presented a minimal model that illustrates the versatility of nonreciprocity as an added dimension to a system’s phase diagram and found that it generates a wealth of stationary and dynamical phase transitions.
This added dimension to the phase diagram allows a simple two-component system to access states of three-fluid coexistence, liquid-liquid phase separation with strong species segregation, and traveling crystal phases.
The minimal system presented here may thus serve as an ideal model for the development of the nonequilibrium statistical mechanics of mixtures.

\textit{Supplemental Material.--} 
See Supplemental Material, which includes Refs.~\cite{Martinez2009, Stukowski2009, Weber2016, Hargus2021, Weeks1971}, for additional simulation and calculation details, as well as simulation movies.

\begin{acknowledgments}
Y.-J.C. acknowledges support from the UC Berkeley College of Engineering Jane Lewis Fellowship. 
A.K.O. was supported by the Laboratory Directed Research and Development Program of Lawrence Berkeley National Laboratory under U.S. Department of Energy Contract No. DE-AC02-05CH11231 and the UC Berkeley College of Engineering.
This research used the Savio computational cluster resource provided by the Berkeley Research Computing program. 
The data that support the findings of this study are available from the corresponding author upon reasonable request.
\end{acknowledgments}

\appendix*
\section{} 
{\color{black} The interaction reciprocity in our simulations is adjusted with a single scalar parameter, $\Delta$.
All interactions within the reciprocity diameter $d_{\rm rec} = 2^{1/6}\sigma$ are reciprocal, and are generally nonreciprocal for interparticle separations greater than $d_{\rm rec}$ for interspecies pairs.
Figure~\ref{fig:forces} plots the pairwise forces resulting from our choice of the reciprocity diameter and the use of a Lennard-Jones potential. 
The sign convention is chosen such that positive forces are repulsive and negative forces are attractive.
The emergence of nonreciprocity is evident with increasing $\Delta$.} 

Particle dynamics are taken to follow the overdamped Langevin equation: 
\begin{equation}
    \label{eq:langevin}
    \dot{\mathbf{x}}_{i} = \frac{1}{\zeta}\sum_{i \neq j}\mathbf{F}_{ij}(r;t) + \mathbf{v}^{s}_{i}(t)
\end{equation}
where $\mathbf{\dot{x}}_i$ is the velocity of the $i$th particle, $\mathbf{v}^{s}_{i}(t)$ is a stochastic velocity with a mean of $\langle\mathbf{v}^{s}_{i}(t) \rangle = \mathbf{0}$ and variance of $\langle \mathbf{v}^{s}_{i}(t)\mathbf{v}^{s}_{j}(t')\rangle = 2D_{T}\delta_{ij}\delta(t-t')\mathbf{I}$, $D_{T}$ is the translational diffusion coefficient (defining the thermal energy scale $k_BT = \zeta D_T$), $\delta(t-t')$ is the Dirac delta function, and $\mathbf{I}$ is the identity tensor.
These conservative forces are cutoff at $2.5\sigma$, capturing the attractive portion of the LJ potential and we set $d_{\rm rec} = 2^{1/6}\sigma$ such that all particle pairs experience reciprocal repulsion within separation distances of $d_{\rm rec}$.
All points sampled in the phase diagram are initialized with random initial conditions generated with \texttt{Packmol}~\cite{Martinez2009} and are run for a duration of $5000\tau$.  
Points close to the phase boundary are examined with different initial conditions to assess global stability, as detailed in the SI~\cite{Note1}.  

{\color{black}Periodic boundary conditions are employed in all directions and rectangular box geometries are used to orient the interface of coexisting domains. 
The relative box dimensions for all simulations have a ratio 1:1:3. 
In the case of traveling states that consist of spatial gradients over large length scales, we conduct additional simulations with relative box dimensions 1:1:10. 
This elongated geometry better allows for visualization and characterization of the spatial gradients present in the traveling states.}

The correlation lengths presented in Fig.~\ref{fig:clustering}(a) are taken to be the inverse of the first moment of the static structure factor:
\begin{equation}
    \label{correlationlength}
    \xi_\alpha = \frac{\int_{k_{\rm min}}^{k^*}S_{\alpha\alpha}(k) dk}{\int_{k_{\rm min}}^{k^*}kS_{\alpha\alpha}(k) dk},
\end{equation}
where $k^*$ is the magnitude of the wavevector corresponding to the reciprocity diameter and $k_{\rm min}$ corresponds to the largest possible wavelength, set by the simulation box size.
The partial static structure factors $S_{\alpha\beta}$ are computed with the Fourier transform of particle coordination for species $\alpha$ and $\beta$. 
We focus on \textit{intraspecies} density correlations (i.e.,~$\alpha=\beta$) with $S_{\alpha\alpha}$ taking the following form:
\begin{equation}
    \label{staticstructurefactor}
    S_{\alpha \alpha}(\mathbf{k}) = \frac{1}{N_{\alpha}}\left\langle\left|\sum_{j = 1}^{N_{\alpha}}\exp(\mathrm{i}\mathbf{k}\cdot\mathbf{x}_{j})\right|^2\right\rangle,
\end{equation}
where $N_{\alpha}$ is the total number of particles of species $\alpha$.
For the isotropic systems considered here, $S_{\alpha \alpha}(k)$ depends only on the magnitude of the wavevector $k = |\mathbf{k}|$.
Species correlation is normalized by $\xi_0 = \frac{\int_{k_{\rm min}}^{k^*}S_{\rm ideal}(k) dk}{\int_{k_{\rm min}}^{k^*}kS_{\rm ideal}(k) dk}$, where $S_{\rm ideal}(k) = 1$ is the ideal gas static structure factor. 

The Steinhardt-Nelson-Ronchetti order parameter $q_{12}$ computed in Fig.~\ref{fig:traveling}(a) is defined as:
\begin{equation}
    \label{steinhardt1}
    q_{l}(i) = \left({\frac{4\pi}{2l+1}\sum_{m = -l}^{l}|\langle q_{lm}\rangle|^2}\right)^{1/2},
\end{equation}
where $q_{lm}$ is the average spherical harmonics of the bond angles formed between particle $i$ and its nearest neighbors~\cite{Steinhardt1983}.
By taking $l = 12$, we can quantify the presence of twelve-fold rotational symmetry in the local structure around a tagged particle with $q_{12} \approx 0.6$ corresponding to a perfect fcc arrangement and $q_{12} \approx 0.3$ for a disordered fluid.
We use $q_{12}$ to distinguish between traveling liquid-gas and traveling crystal-gas scenarios.

\begin{thebibliography}{66}%
\makeatletter
\providecommand \@ifxundefined [1]{%
 \@ifx{#1\undefined}
}%
\providecommand \@ifnum [1]{%
 \ifnum #1\expandafter \@firstoftwo
 \else \expandafter \@secondoftwo
 \fi
}%
\providecommand \@ifx [1]{%
 \ifx #1\expandafter \@firstoftwo
 \else \expandafter \@secondoftwo
 \fi
}%
\providecommand \natexlab [1]{#1}%
\providecommand \enquote  [1]{``#1''}%
\providecommand \bibnamefont  [1]{#1}%
\providecommand \bibfnamefont [1]{#1}%
\providecommand \citenamefont [1]{#1}%
\providecommand \href@noop [0]{\@secondoftwo}%
\providecommand \href [0]{\begingroup \@sanitize@url \@href}%
\providecommand \@href[1]{\@@startlink{#1}\@@href}%
\providecommand \@@href[1]{\endgroup#1\@@endlink}%
\providecommand \@sanitize@url [0]{\catcode `\\12\catcode `\$12\catcode
  `\&12\catcode `\#12\catcode `\^12\catcode `\_12\catcode `\%12\relax}%
\providecommand \@@startlink[1]{}%
\providecommand \@@endlink[0]{}%
\providecommand \url  [0]{\begingroup\@sanitize@url \@url }%
\providecommand \@url [1]{\endgroup\@href {#1}{\urlprefix }}%
\providecommand \urlprefix  [0]{URL }%
\providecommand \Eprint [0]{\href }%
\providecommand \doibase [0]{https://doi.org/}%
\providecommand \selectlanguage [0]{\@gobble}%
\providecommand \bibinfo  [0]{\@secondoftwo}%
\providecommand \bibfield  [0]{\@secondoftwo}%
\providecommand \translation [1]{[#1]}%
\providecommand \BibitemOpen [0]{}%
\providecommand \bibitemStop [0]{}%
\providecommand \bibitemNoStop [0]{.\EOS\space}%
\providecommand \EOS [0]{\spacefactor3000\relax}%
\providecommand \BibitemShut  [1]{\csname bibitem#1\endcsname}%
\let\auto@bib@innerbib\@empty
\bibitem [{\citenamefont {Long}\ and\ \citenamefont {Azam}(2001)}]{Long2001}%
  \BibitemOpen
  \bibfield  {author} {\bibinfo {author} {\bibfnamefont {R.~A.}\ \bibnamefont
  {Long}}\ and\ \bibinfo {author} {\bibfnamefont {F.}~\bibnamefont {Azam}},\
  }\href {https://journals.asm.org/doi/abs/10.1128/AEM.67.11.4975-4983.2001}
  {\bibfield  {journal} {\bibinfo  {journal} {Appl. Environ. Microbiol.}\ }\textbf
  {\bibinfo {volume} {67}},\ \bibinfo {pages} {4975} (\bibinfo {year}
  {2001})}\BibitemShut {NoStop}%
\bibitem [{\citenamefont {Czir{\'{o}}k}\ \emph {et~al.}(1996)\citenamefont
  {Czir{\'{o}}k}, \citenamefont {Ben-Jacob}, \citenamefont {Cohen},\ and\
  \citenamefont {Vicsek}}]{Czirok1996}%
  \BibitemOpen
  \bibfield  {author} {\bibinfo {author} {\bibfnamefont {A.}~\bibnamefont
  {Czir{\'{o}}k}}, \bibinfo {author} {\bibfnamefont {E.}~\bibnamefont
  {Ben-Jacob}}, \bibinfo {author} {\bibfnamefont {I.}~\bibnamefont {Cohen}},\
  and\ \bibinfo {author} {\bibfnamefont {T.}~\bibnamefont {Vicsek}},\ }\href
  {https://doi.org/10.1103/PhysRevE.54.1791} {\bibfield  {journal} {\bibinfo
  {journal} {Phys. Rev. E}\ }\textbf {\bibinfo {volume} {54}},\ \bibinfo
  {pages} {1791} (\bibinfo {year} {1996})}\BibitemShut {NoStop}%
\bibitem [{\citenamefont {Vicsek}\ and\ \citenamefont
  {Zafeiris}(2012)}]{Vicsek2012}%
  \BibitemOpen
  \bibfield  {author} {\bibinfo {author} {\bibfnamefont {T.}~\bibnamefont
  {Vicsek}}\ and\ \bibinfo {author} {\bibfnamefont {A.}~\bibnamefont
  {Zafeiris}},\ }\href {https://doi.org/10.1016/J.PHYSREP.2012.03.004}
  {\bibfield  {journal} {\bibinfo  {journal} {Phys. Rep.}\ }\textbf {\bibinfo
  {volume} {517}},\ \bibinfo {pages} {71} (\bibinfo {year} {2012})}\BibitemShut
  {NoStop}%
\bibitem [{\citenamefont {Strandburg-Peshkin}\ \emph
  {et~al.}(2013)\citenamefont {Strandburg-Peshkin}, \citenamefont {Twomey},
  \citenamefont {Bode}, \citenamefont {Kao}, \citenamefont {Katz},
  \citenamefont {Ioannou}, \citenamefont {Rosenthal}, \citenamefont {Torney},
  \citenamefont {Wu}, \citenamefont {Levin},\ and\ \citenamefont
  {Couzin}}]{Strandburg2013}%
  \BibitemOpen
  \bibfield  {author} {\bibinfo {author} {\bibfnamefont {A.}~\bibnamefont
  {Strandburg-Peshkin}}, \bibinfo {author} {\bibfnamefont {C.~R.}\ \bibnamefont
  {Twomey}}, \bibinfo {author} {\bibfnamefont {N.~W.~F.}\ \bibnamefont {Bode}},
  \bibinfo {author} {\bibfnamefont {A.~B.}\ \bibnamefont {Kao}}, \bibinfo
  {author} {\bibfnamefont {Y.}~\bibnamefont {Katz}}, \bibinfo {author}
  {\bibfnamefont {C.~C.}\ \bibnamefont {Ioannou}}, \bibinfo {author}
  {\bibfnamefont {S.~B.}\ \bibnamefont {Rosenthal}}, \bibinfo {author}
  {\bibfnamefont {C.~J.}\ \bibnamefont {Torney}}, \bibinfo {author}
  {\bibfnamefont {H.~S.}\ \bibnamefont {Wu}}, \bibinfo {author} {\bibfnamefont
  {S.~A.}\ \bibnamefont {Levin}},\ and\ \bibinfo {author} {\bibfnamefont
  {I.~D.}\ \bibnamefont {Couzin}},\ }\href
  {https://doi.org/10.1016/J.CUB.2013.07.059} {\bibfield  {journal} {\bibinfo
  {journal} {Curr. Biol.}\ }\textbf {\bibinfo {volume} {23}},\ \bibinfo {pages}
  {R709} (\bibinfo {year} {2013})}\BibitemShut {NoStop}%
\bibitem [{\citenamefont {Tsytovich}(1997)}]{Tsytovich1997}%
  \BibitemOpen
  \bibfield  {author} {\bibinfo {author} {\bibfnamefont {V.~N.}\ \bibnamefont
  {Tsytovich}},\ }\href {https://doi.org/10.1070/PU1997v040n01ABEH000201}
  {\bibfield  {journal} {\bibinfo  {journal} {Phys.-Usp.}\ }\textbf {\bibinfo
  {volume} {40}},\ \bibinfo {pages} {53} (\bibinfo {year} {1997})}\BibitemShut
  {NoStop}%
\bibitem [{\citenamefont {Chaudhuri}\ \emph {et~al.}(2011)\citenamefont
  {Chaudhuri}, \citenamefont {Ivlev}, \citenamefont {Khrapak}, \citenamefont
  {Thomas},\ and\ \citenamefont {Morfill}}]{Chaudhuri2011}%
  \BibitemOpen
  \bibfield  {author} {\bibinfo {author} {\bibfnamefont {M.}~\bibnamefont
  {Chaudhuri}}, \bibinfo {author} {\bibfnamefont {A.~V.}\ \bibnamefont
  {Ivlev}}, \bibinfo {author} {\bibfnamefont {S.~A.}\ \bibnamefont {Khrapak}},
  \bibinfo {author} {\bibfnamefont {H.~M.}\ \bibnamefont {Thomas}},\ and\
  \bibinfo {author} {\bibfnamefont {G.~E.}\ \bibnamefont {Morfill}},\ }\href
  {https://doi.org/10.1039/C0SM00813C} {\bibfield  {journal} {\bibinfo
  {journal} {Soft Matter}\ }\textbf {\bibinfo {volume} {7}},\ \bibinfo {pages}
  {1287} (\bibinfo {year} {2011})}\BibitemShut {NoStop}%
\bibitem [{\citenamefont {Morfill}\ and\ \citenamefont
  {Ivlev}(2009)}]{Morfill2009}%
  \BibitemOpen
  \bibfield  {author} {\bibinfo {author} {\bibfnamefont {G.~E.}\ \bibnamefont
  {Morfill}}\ and\ \bibinfo {author} {\bibfnamefont {A.~V.}\ \bibnamefont
  {Ivlev}},\ }\href {https://link.aps.org/doi/10.1103/RevModPhys.81.1353}
  {\bibfield  {journal} {\bibinfo  {journal} {Rev. Mod. Phys.}\ }\textbf
  {\bibinfo {volume} {81}},\ \bibinfo {pages} {1353} (\bibinfo {year}
  {2009})}\BibitemShut {NoStop}%
\bibitem [{\citenamefont {Keh}(2016)}]{Keh2016}%
  \BibitemOpen
  \bibfield  {author} {\bibinfo {author} {\bibfnamefont {H.~J.}\ \bibnamefont
  {Keh}},\ }\href {https://doi.org/10.1016/J.COCIS.2016.05.008} {\bibfield
  {journal} {\bibinfo  {journal} {Curr. Opin. Colloid Interface Sci.}\ }\textbf {\bibinfo
  {volume} {24}},\ \bibinfo {pages} {13} (\bibinfo {year} {2016})}\BibitemShut
  {NoStop}%
\bibitem [{\citenamefont {Ivlev}\ and\ \citenamefont
  {Kompaneets}(2017)}]{Ivlev2017}%
  \BibitemOpen
  \bibfield  {author} {\bibinfo {author} {\bibfnamefont {A.~V.}\ \bibnamefont
  {Ivlev}}\ and\ \bibinfo {author} {\bibfnamefont {R.}~\bibnamefont
  {Kompaneets}},\ }\href
  {https://journals.aps.org/pre/abstract/10.1103/PhysRevE.95.053202} {\bibfield
   {journal} {\bibinfo  {journal} {Phys. Rev. E}\ }\textbf {\bibinfo {volume}
  {95}},\ \bibinfo {pages} {053202} (\bibinfo {year} {2017})}\BibitemShut
  {NoStop}%
\bibitem [{\citenamefont {Kompaneets}\ \emph {et~al.}(2016)\citenamefont
  {Kompaneets}, \citenamefont {Morfill},\ and\ \citenamefont
  {Ivlev}}]{Kompaneets2016}%
  \BibitemOpen
  \bibfield  {author} {\bibinfo {author} {\bibfnamefont {R.}~\bibnamefont
  {Kompaneets}}, \bibinfo {author} {\bibfnamefont {G.~E.}\ \bibnamefont
  {Morfill}},\ and\ \bibinfo {author} {\bibfnamefont {A.~V.}\ \bibnamefont
  {Ivlev}},\ }\href
  {https://journals.aps.org/pre/abstract/10.1103/PhysRevE.93.063201} {\bibfield
   {journal} {\bibinfo  {journal} {Phys. Rev. E}\ }\textbf {\bibinfo {volume}
  {93}},\ \bibinfo {pages} {063201} (\bibinfo {year} {2016})}\BibitemShut
  {NoStop}%
\bibitem [{\citenamefont {Meredith}\ \emph {et~al.}(2020)\citenamefont
  {Meredith}, \citenamefont {Moerman}, \citenamefont {Groenewold},
  \citenamefont {Chiu}, \citenamefont {Kegel}, \citenamefont {van Blaaderen},\
  and\ \citenamefont {Zarzar}}]{Meredith2020}%
  \BibitemOpen
  \bibfield  {author} {\bibinfo {author} {\bibfnamefont {C.~H.}\ \bibnamefont
  {Meredith}}, \bibinfo {author} {\bibfnamefont {P.~G.}\ \bibnamefont
  {Moerman}}, \bibinfo {author} {\bibfnamefont {J.}~\bibnamefont {Groenewold}},
  \bibinfo {author} {\bibfnamefont {Y.-J.}\ \bibnamefont {Chiu}}, \bibinfo
  {author} {\bibfnamefont {W.~K.}\ \bibnamefont {Kegel}}, \bibinfo {author}
  {\bibfnamefont {A.}~\bibnamefont {van Blaaderen}},\ and\ \bibinfo {author}
  {\bibfnamefont {L.~D.}\ \bibnamefont {Zarzar}},\ }\href
  {https://doi.org/10.1038/s41557-020-00575-0} {\bibfield  {journal} {\bibinfo
  {journal} {Nat. Chem.}\ }\textbf {\bibinfo {volume} {12}},\ \bibinfo {pages}
  {1136} (\bibinfo {year} {2020})}\BibitemShut {NoStop}%
\bibitem [{\citenamefont {Sriram}\ and\ \citenamefont
  {Furst}(2012)}]{Sriram2012}%
  \BibitemOpen
  \bibfield  {author} {\bibinfo {author} {\bibfnamefont {I.}~\bibnamefont
  {Sriram}}\ and\ \bibinfo {author} {\bibfnamefont {E.~M.}\ \bibnamefont
  {Furst}},\ }\href {https://doi.org/10.1039/C2SM06784F} {\bibfield  {journal}
  {\bibinfo  {journal} {Soft Matter}\ }\textbf {\bibinfo {volume} {8}},\
  \bibinfo {pages} {3335} (\bibinfo {year} {2012})}\BibitemShut {NoStop}%
\bibitem [{\citenamefont {Soto}\ and\ \citenamefont
  {Golestanian}(2014)}]{Soto2014}%
  \BibitemOpen
  \bibfield  {author} {\bibinfo {author} {\bibfnamefont {R.}~\bibnamefont
  {Soto}}\ and\ \bibinfo {author} {\bibfnamefont {R.}~\bibnamefont
  {Golestanian}},\ }\href {https://doi.org/10.1103/PHYSREVLETT.112.068301}
  {\bibfield  {journal} {\bibinfo  {journal} {Phys. Rev. Lett.}\ }\textbf
  {\bibinfo {volume} {112}},\ \bibinfo {pages} {068301} (\bibinfo {year}
  {2014})}\BibitemShut {NoStop}%
\bibitem [{\citenamefont {Saha}\ \emph {et~al.}(2019)\citenamefont {Saha},
  \citenamefont {Ramaswamy},\ and\ \citenamefont {Golestanian}}]{Saha2019}%
  \BibitemOpen
  \bibfield  {author} {\bibinfo {author} {\bibfnamefont {S.}~\bibnamefont
  {Saha}}, \bibinfo {author} {\bibfnamefont {S.}~\bibnamefont {Ramaswamy}},\
  and\ \bibinfo {author} {\bibfnamefont {R.}~\bibnamefont {Golestanian}},\
  }\href {https://doi.org/10.1088/1367-2630/AB20FD} {\bibfield  {journal}
  {\bibinfo  {journal} {New J. Phys.}\ }\textbf {\bibinfo {volume} {21}},\
  \bibinfo {pages} {063006} (\bibinfo {year} {2019})}\BibitemShut {NoStop}%
\bibitem [{\citenamefont {Scheibner}\ \emph {et~al.}(2020)\citenamefont
  {Scheibner}, \citenamefont {Souslov}, \citenamefont {Banerjee}, \citenamefont
  {Sur{\'{o}}wka}, \citenamefont {Irvine},\ and\ \citenamefont
  {Vitelli}}]{Scheibner2020}%
  \BibitemOpen
  \bibfield  {author} {\bibinfo {author} {\bibfnamefont {C.}~\bibnamefont
  {Scheibner}}, \bibinfo {author} {\bibfnamefont {A.}~\bibnamefont {Souslov}},
  \bibinfo {author} {\bibfnamefont {D.}~\bibnamefont {Banerjee}}, \bibinfo
  {author} {\bibfnamefont {P.}~\bibnamefont {Sur{\'{o}}wka}}, \bibinfo {author}
  {\bibfnamefont {W.~T.~M.}\ \bibnamefont {Irvine}},\ and\ \bibinfo {author}
  {\bibfnamefont {V.}~\bibnamefont {Vitelli}},\ }\href
  {https://doi.org/10.1038/s41567-020-0795-y} {\bibfield  {journal} {\bibinfo
  {journal} {Nat. Phys.}\ }\textbf {\bibinfo {volume} {16}},\ \bibinfo {pages}
  {475} (\bibinfo {year} {2020})}\BibitemShut {NoStop}%
\bibitem [{\citenamefont {Miri}\ and\ \citenamefont
  {Al{\`{u}}}(2019)}]{Miri2019}%
  \BibitemOpen
  \bibfield  {author} {\bibinfo {author} {\bibfnamefont {M.~A.}\ \bibnamefont
  {Miri}}\ and\ \bibinfo {author} {\bibfnamefont {A.}~\bibnamefont
  {Al{\`{u}}}},\ }\href {https://www.science.org/doi/10.1126/science.aar7709}
  {\bibfield  {journal} {\bibinfo  {journal} {Science}\ }\textbf {\bibinfo
  {volume} {363}} \bibinfo{page} {eaar7709} (\bibinfo {year} {2019})}\BibitemShut {NoStop}%
\bibitem [{\citenamefont {Nassar}\ \emph {et~al.}(2020)\citenamefont {Nassar},
  \citenamefont {Yousefzadeh}, \citenamefont {Fleury}, \citenamefont {Ruzzene},
  \citenamefont {Al{\`{u}}}, \citenamefont {Daraio}, \citenamefont {Norris},
  \citenamefont {Huang},\ and\ \citenamefont {Haberman}}]{Nassar2020}%
  \BibitemOpen
  \bibfield  {author} {\bibinfo {author} {\bibfnamefont {H.}~\bibnamefont
  {Nassar}}, \bibinfo {author} {\bibfnamefont {B.}~\bibnamefont {Yousefzadeh}},
  \bibinfo {author} {\bibfnamefont {R.}~\bibnamefont {Fleury}}, \bibinfo
  {author} {\bibfnamefont {M.}~\bibnamefont {Ruzzene}}, \bibinfo {author}
  {\bibfnamefont {A.}~\bibnamefont {Al{\`{u}}}}, \bibinfo {author}
  {\bibfnamefont {C.}~\bibnamefont {Daraio}}, \bibinfo {author} {\bibfnamefont
  {A.~N.}\ \bibnamefont {Norris}}, \bibinfo {author} {\bibfnamefont
  {G.}~\bibnamefont {Huang}},\ and\ \bibinfo {author} {\bibfnamefont {M.~R.}\
  \bibnamefont {Haberman}},\ }\href {https://doi.org/10.1038/s41578-020-0206-0}
  {\bibfield  {journal} {\bibinfo  {journal} {Nat. Rev. Mater.}\ }\textbf
  {\bibinfo {volume} {5}},\ \bibinfo {pages} {667} (\bibinfo {year}
  {2020})}\BibitemShut {NoStop}%
\bibitem [{\citenamefont {Israelachvili}(1992)}]{Israelachvili1992}%
  \BibitemOpen
  \bibfield  {author} {\bibinfo {author} {\bibfnamefont {J.~N.}\ \bibnamefont
  {Israelachvili}},\ }\href
  {https://doi.org/10.1016/B978-0-12-375182-9.10001-6} {\emph {\bibinfo {title}
  {{Intermolecular and Surface Forces}}}}\ (\bibinfo  {publisher} {Elsevier},\
  \bibinfo {address} {Amsterdam},\ \bibinfo {year} {1992})\BibitemShut
  {NoStop}%
\bibitem [{\citenamefont {Dijkstra}\ \emph {et~al.}(2000)\citenamefont
  {Dijkstra}, \citenamefont {Van~Roij},\ and\ \citenamefont
  {Evans}}]{Dijkstra2000}%
  \BibitemOpen
  \bibfield  {author} {\bibinfo {author} {\bibfnamefont {M.}~\bibnamefont
  {Dijkstra}}, \bibinfo {author} {\bibfnamefont {R.}~\bibnamefont {van~Roij}},\
  and\ \bibinfo {author} {\bibfnamefont {R.}~\bibnamefont {Evans}},\ }\href
  {https://doi.org/10.1063/1.1288921} {\bibfield  {journal} {\bibinfo
  {journal} {J. Chem. Phys.}\ }\textbf {\bibinfo {volume} {113}},\ \bibinfo
  {pages} {4799} (\bibinfo {year} {2000})}\BibitemShut {NoStop}%
\bibitem [{\citenamefont {Bolhuis}\ \emph {et~al.}(2001)\citenamefont
  {Bolhuis}, \citenamefont {Louis}, \citenamefont {Hansen},\ and\ \citenamefont
  {Meijer}}]{Bolhuis2001}%
  \BibitemOpen
  \bibfield  {author} {\bibinfo {author} {\bibfnamefont {P.~G.}\ \bibnamefont
  {Bolhuis}}, \bibinfo {author} {\bibfnamefont {A.~A.}\ \bibnamefont {Louis}},
  \bibinfo {author} {\bibfnamefont {J.~P.}\ \bibnamefont {Hansen}},\ and\
  \bibinfo {author} {\bibfnamefont {E.~J.}\ \bibnamefont {Meijer}},\ }\href
  {https://doi.org/10.1063/1.1344606} {\bibfield  {journal} {\bibinfo
  {journal} {J. Chem. Phys.}\ }\textbf {\bibinfo {volume} {114}},\ \bibinfo
  {pages} {4296} (\bibinfo {year} {2001})}\BibitemShut {NoStop}%
\bibitem [{\citenamefont {Praprotnik}\ \emph {et~al.}(2008)\citenamefont
  {Praprotnik}, \citenamefont {Site},\ and\ \citenamefont
  {Kremer}}]{Praprotnik2008}%
  \BibitemOpen
  \bibfield  {author} {\bibinfo {author} {\bibfnamefont {M.}~\bibnamefont
  {Praprotnik}}, \bibinfo {author} {\bibfnamefont {L.~D.}\ \bibnamefont
  {Site}},\ and\ \bibinfo {author} {\bibfnamefont {K.}~\bibnamefont {Kremer}},\
  }\href {https://doi.org/10.1146/annurev.physchem.59.032607.093707} {\bibfield
   {journal} {\bibinfo  {journal} {Annu. Rev. Phys. Chem.}\ }\textbf {\bibinfo
  {volume} {59}},\ \bibinfo {pages} {545} (\bibinfo {year} {2008})}\BibitemShut
  {NoStop}%
\bibitem [{\citenamefont {Mognetti}\ \emph {et~al.}(2009)\citenamefont
  {Mognetti}, \citenamefont {Virnau}, \citenamefont {Yelash}, \citenamefont
  {Paul}, \citenamefont {Binder}, \citenamefont {M{\"{u}}ller},\ and\
  \citenamefont {MacDowell}}]{Mognetti2009}%
  \BibitemOpen
  \bibfield  {author} {\bibinfo {author} {\bibfnamefont {B.~M.}\ \bibnamefont
  {Mognetti}}, \bibinfo {author} {\bibfnamefont {P.}~\bibnamefont {Virnau}},
  \bibinfo {author} {\bibfnamefont {L.}~\bibnamefont {Yelash}}, \bibinfo
  {author} {\bibfnamefont {W.}~\bibnamefont {Paul}}, \bibinfo {author}
  {\bibfnamefont {K.}~\bibnamefont {Binder}}, \bibinfo {author} {\bibfnamefont
  {M.}~\bibnamefont {M{\"{u}}ller}},\ and\ \bibinfo {author} {\bibfnamefont
  {L.~G.}\ \bibnamefont {MacDowell}},\ }\href
  {https://doi.org/10.1063/1.3050353} {\bibfield  {journal} {\bibinfo
  {journal} {J. Chem. Phys.}\ }\textbf {\bibinfo {volume} {130}},\ \bibinfo
  {pages} {044101} (\bibinfo {year} {2009})}\BibitemShut {NoStop}%
\bibitem [{\citenamefont {Cates}\ and\ \citenamefont
  {Tailleur}(2015)}]{Cates2015}%
  \BibitemOpen
  \bibfield  {author} {\bibinfo {author} {\bibfnamefont {M.~E.}\ \bibnamefont
  {Cates}}\ and\ \bibinfo {author} {\bibfnamefont {J.}~\bibnamefont
  {Tailleur}},\ }\href
  {https://www.annualreviews.org/doi/abs/10.1146/annurev-conmatphys-031214-014710}
  {\bibfield  {journal} {\bibinfo  {journal} {Annu. Rev. Condens. Matter
  Phys.}\ }\textbf {\bibinfo {volume} {6}},\ \bibinfo {pages} {219} (\bibinfo
  {year} {2015})}\BibitemShut {NoStop}%
\bibitem [{\citenamefont {Bechinger}\ \emph {et~al.}(2016)\citenamefont
  {Bechinger}, \citenamefont {Di~Leonardo}, \citenamefont {L{\"{o}}wen},
  \citenamefont {Reichhardt}, \citenamefont {Volpe},\ and\ \citenamefont
  {Volpe}}]{Bechinger2016}%
  \BibitemOpen
  \bibfield  {author} {\bibinfo {author} {\bibfnamefont {C.}~\bibnamefont
  {Bechinger}}, \bibinfo {author} {\bibfnamefont {R.}~\bibnamefont
  {Di~Leonardo}}, \bibinfo {author} {\bibfnamefont {H.}~\bibnamefont
  {L{\"{o}}wen}}, \bibinfo {author} {\bibfnamefont {C.}~\bibnamefont
  {Reichhardt}}, \bibinfo {author} {\bibfnamefont {G.}~\bibnamefont {Volpe}},\
  and\ \bibinfo {author} {\bibfnamefont {G.}~\bibnamefont {Volpe}},\ }\href
  {https://journals.aps.org/rmp/abstract/10.1103/RevModPhys.88.045006}
  {\bibfield  {journal} {\bibinfo  {journal} {Rev. Mod. Phys.}\ }\textbf
  {\bibinfo {volume} {88}},\ \bibinfo {pages} {045006} (\bibinfo {year}
  {2016})}\BibitemShut {NoStop}%
\bibitem [{\citenamefont {Hayashi}\ and\ \citenamefont
  {Sasa}(2006)}]{Hayashi2006}%
  \BibitemOpen
  \bibfield  {author} {\bibinfo {author} {\bibfnamefont {K.}~\bibnamefont
  {Hayashi}}\ and\ \bibinfo {author} {\bibfnamefont {S.-i.}\ \bibnamefont
  {Sasa}},\ }\href {https://doi.org/10.1088/0953-8984/18/10/008} {\bibfield
  {journal} {\bibinfo  {journal} {J. Phys.: Condens. Matter.}\ }\textbf
  {\bibinfo {volume} {18}},\ \bibinfo {pages} {2825} (\bibinfo {year}
  {2006})}\BibitemShut {NoStop}%
\bibitem [{\citenamefont {Durve}\ \emph {et~al.}(2018)\citenamefont {Durve},
  \citenamefont {Saha},\ and\ \citenamefont {Sayeed}}]{Durve2018}%
  \BibitemOpen
  \bibfield  {author} {\bibinfo {author} {\bibfnamefont {M.}~\bibnamefont
  {Durve}}, \bibinfo {author} {\bibfnamefont {A.}~\bibnamefont {Saha}},\ and\
  \bibinfo {author} {\bibfnamefont {A.}~\bibnamefont {Sayeed}},\ }\href
  {https://doi.org/10.1140/EPJE/I2018-11653-4} {\bibfield  {journal} {\bibinfo
  {journal} {Eur. Phys. J. E}\ }\textbf {\bibinfo {volume} {49}},\ \bibinfo
  {pages} {1} (\bibinfo {year} {2018})}\BibitemShut {NoStop}%
\bibitem [{\citenamefont {Ivlev}\ \emph {et~al.}(2015)\citenamefont {Ivlev},
  \citenamefont {Bartnick}, \citenamefont {Heinen}, \citenamefont {Du},
  \citenamefont {Nosenko},\ and\ \citenamefont {L{\"{o}}wen}}]{Ivlev2015}%
  \BibitemOpen
  \bibfield  {author} {\bibinfo {author} {\bibfnamefont {A.~V.}\ \bibnamefont
  {Ivlev}}, \bibinfo {author} {\bibfnamefont {J.}~\bibnamefont {Bartnick}},
  \bibinfo {author} {\bibfnamefont {M.}~\bibnamefont {Heinen}}, \bibinfo
  {author} {\bibfnamefont {C.~R.}\ \bibnamefont {Du}}, \bibinfo {author}
  {\bibfnamefont {V.}~\bibnamefont {Nosenko}},\ and\ \bibinfo {author}
  {\bibfnamefont {H.}~\bibnamefont {L{\"{o}}wen}},\ }\href
  {https://doi.org/10.1103/PHYSREVX.5.011035} {\bibfield  {journal} {\bibinfo
  {journal} {Phys. Rev. X}\ }\textbf {\bibinfo {volume} {5}},\ \bibinfo {pages}
  {011035} (\bibinfo {year} {2015})}\BibitemShut {NoStop}%
\bibitem [{\citenamefont {Agudo-Canalejo}\ and\ \citenamefont
  {Golestanian}(2019)}]{Agudo2019}%
  \BibitemOpen
  \bibfield  {author} {\bibinfo {author} {\bibfnamefont {J.}~\bibnamefont
  {Agudo-Canalejo}}\ and\ \bibinfo {author} {\bibfnamefont {R.}~\bibnamefont
  {Golestanian}},\ }\href
  {https://doi.org/https://doi.org/10.1103/PhysRevLett.123.018101} {\bibfield
  {journal} {\bibinfo  {journal} {Phys. Rev. Lett.}\ }\textbf {\bibinfo
  {volume} {123}},\ \bibinfo {pages} {018101} (\bibinfo {year}
  {2019})}\BibitemShut {NoStop}%
\bibitem [{\citenamefont {You}\ \emph {et~al.}(2020)\citenamefont {You},
  \citenamefont {Baskaran},\ and\ \citenamefont {Marchetti}}]{You2020}%
  \BibitemOpen
  \bibfield  {author} {\bibinfo {author} {\bibfnamefont {Z.}~\bibnamefont
  {You}}, \bibinfo {author} {\bibfnamefont {A.}~\bibnamefont {Baskaran}},\ and\
  \bibinfo {author} {\bibfnamefont {M.~C.}\ \bibnamefont {Marchetti}},\ }\href
  {https://doi.org/10.1073/pnas.2010318117} {\bibfield  {journal} {\bibinfo
  {journal} {Proc. Natl. Acad. Sci. USA}\ }\textbf {\bibinfo {volume} {117}},\
  \bibinfo {pages} {19767} (\bibinfo {year} {2020})}\BibitemShut {NoStop}%
\bibitem [{\citenamefont {Saha}\ \emph {et~al.}(2020)\citenamefont {Saha},
  \citenamefont {Agudo-Canalejo},\ and\ \citenamefont
  {Golestanian}}]{Saha2020}%
  \BibitemOpen
  \bibfield  {author} {\bibinfo {author} {\bibfnamefont {S.}~\bibnamefont
  {Saha}}, \bibinfo {author} {\bibfnamefont {J.}~\bibnamefont
  {Agudo-Canalejo}},\ and\ \bibinfo {author} {\bibfnamefont {R.}~\bibnamefont
  {Golestanian}},\ }\href {https://doi.org/10.1103/PhysRevX.10.041009}
  {\bibfield  {journal} {\bibinfo  {journal} {Phys. Rev. X}\ }\textbf {\bibinfo
  {volume} {10}},\ \bibinfo {pages} {041009} (\bibinfo {year}
  {2020})}\BibitemShut {NoStop}%
\bibitem [{\citenamefont {Fruchart}\ \emph {et~al.}(2021)\citenamefont
  {Fruchart}, \citenamefont {Hanai}, \citenamefont {Littlewood},\ and\
  \citenamefont {Vitelli}}]{Fruchart2021}%
  \BibitemOpen
  \bibfield  {author} {\bibinfo {author} {\bibfnamefont {M.}~\bibnamefont
  {Fruchart}}, \bibinfo {author} {\bibfnamefont {R.}~\bibnamefont {Hanai}},
  \bibinfo {author} {\bibfnamefont {P.~B.}\ \bibnamefont {Littlewood}},\ and\
  \bibinfo {author} {\bibfnamefont {V.}~\bibnamefont {Vitelli}},\ }\href
  {https://doi.org/10.1038/s41586-021-03375-9} {\bibfield  {journal} {\bibinfo
  {journal} {Nature}\ }\textbf {\bibinfo {volume} {592}},\ \bibinfo {pages}
  {363} (\bibinfo {year} {2021})}\BibitemShut {NoStop}%
\bibitem [{\citenamefont {Saha}\ and\ \citenamefont
  {Golestanian}(2022)}]{Saha2022}%
  \BibitemOpen
  \bibfield  {author} {\bibinfo {author} {\bibfnamefont {S.}~\bibnamefont
  {Saha}}\ and\ \bibinfo {author} {\bibfnamefont {R.}~\bibnamefont
  {Golestanian}},\ }\href {https://arxiv.org/abs/2208.14985v1} {\bibfield
  {journal} {\bibinfo  {journal} {arXiv:2208.14985}\ } (\bibinfo {year}
  {2022})}\BibitemShut {NoStop}%
\bibitem [{\citenamefont {Osat}\ and\ \citenamefont
  {Golestanian}(2022)}]{Osat2022}%
  \BibitemOpen
  \bibfield  {author} {\bibinfo {author} {\bibfnamefont {S.}~\bibnamefont
  {Osat}}\ and\ \bibinfo {author} {\bibfnamefont {R.}~\bibnamefont
  {Golestanian}},\ }\href {https://doi.org/10.1038/s41565-022-01258-2}
  {\bibfield  {journal} {\bibinfo  {journal} {Nat. Nanotechnol.}\ }\textbf
  {\bibinfo {volume} {18}},\ \bibinfo {pages} {79} (\bibinfo {year}
  {2022})}\BibitemShut {NoStop}%
\bibitem [{\citenamefont {Dinelli}\ \emph {et~al.}(2022)\citenamefont
  {Dinelli}, \citenamefont {O'Byrne}, \citenamefont {Curatolo}, \citenamefont
  {Zhao}, \citenamefont {Sollich},\ and\ \citenamefont
  {Tailleur}}]{Dinelli2022}%
  \BibitemOpen
  \bibfield  {author} {\bibinfo {author} {\bibfnamefont {A.}~\bibnamefont
  {Dinelli}}, \bibinfo {author} {\bibfnamefont {J.}~\bibnamefont {O'Byrne}},
  \bibinfo {author} {\bibfnamefont {A.}~\bibnamefont {Curatolo}}, \bibinfo
  {author} {\bibfnamefont {Y.}~\bibnamefont {Zhao}}, \bibinfo {author}
  {\bibfnamefont {P.}~\bibnamefont {Sollich}},\ and\ \bibinfo {author}
  {\bibfnamefont {J.}~\bibnamefont {Tailleur}},\ }\href
  {https://arxiv.org/abs/2212.03995} {\bibfield  {journal} {\bibinfo  {journal}
  {arXiv:2203.07757}\ } (\bibinfo {year} {2022})}\BibitemShut {NoStop}%
\bibitem [{\citenamefont {Bartnick}\ \emph {et~al.}(2015)\citenamefont
  {Bartnick}, \citenamefont {Heinen}, \citenamefont {Ivlev},\ and\
  \citenamefont {L{\"{o}}wen}}]{Bartnick2015}%
  \BibitemOpen
  \bibfield  {author} {\bibinfo {author} {\bibfnamefont {J.}~\bibnamefont
  {Bartnick}}, \bibinfo {author} {\bibfnamefont {M.}~\bibnamefont {Heinen}},
  \bibinfo {author} {\bibfnamefont {A.~V.}\ \bibnamefont {Ivlev}},\ and\
  \bibinfo {author} {\bibfnamefont {H.}~\bibnamefont {L{\"{o}}wen}},\ }\href
  {https://doi.org/10.1088/0953-8984/28/2/025102} {\bibfield  {journal}
  {\bibinfo  {journal} {J. Phys. Condens. Matter}\ }\textbf {\bibinfo {volume}
  {28}},\ \bibinfo {pages} {025102} (\bibinfo {year} {2015})}\BibitemShut
  {NoStop}%
\bibitem [{\citenamefont {Kryuchkov}\ \emph {et~al.}(2018)\citenamefont
  {Kryuchkov}, \citenamefont {Ivlev},\ and\ \citenamefont
  {Yurchenko}}]{Kryuchkov2018}%
  \BibitemOpen
  \bibfield  {author} {\bibinfo {author} {\bibfnamefont {N.~P.}\ \bibnamefont
  {Kryuchkov}}, \bibinfo {author} {\bibfnamefont {A.~V.}\ \bibnamefont
  {Ivlev}},\ and\ \bibinfo {author} {\bibfnamefont {S.~O.}\ \bibnamefont
  {Yurchenko}},\ }\href {https://doi.org/10.1039/C8SM01836G} {\bibfield
  {journal} {\bibinfo  {journal} {Soft Matter}\ }\textbf {\bibinfo {volume}
  {14}},\ \bibinfo {pages} {9720} (\bibinfo {year} {2018})}\BibitemShut
  {NoStop}%
\bibitem [{\citenamefont {Omar}\ \emph {et~al.}(2022)\citenamefont {Omar},
  \citenamefont {Row}, \citenamefont {Mallory},\ and\ \citenamefont
  {Brady}}]{Omar2022}%
  \BibitemOpen
  \bibfield  {author} {\bibinfo {author} {\bibfnamefont {A.~K.}\ \bibnamefont
  {Omar}}, \bibinfo {author} {\bibfnamefont {H.}~\bibnamefont {Row}}, \bibinfo
  {author} {\bibfnamefont {S.~A.}\ \bibnamefont {Mallory}},\ and\ \bibinfo
  {author} {\bibfnamefont {J.~F.}\ \bibnamefont {Brady}},\ }\href
  {https://arxiv.org/abs/2211.12673v1} {\bibfield  {journal} {\bibinfo
  {journal} {arXiv:2211.12673}\ } (\bibinfo {year} {2022})}\BibitemShut
  {NoStop}%
\bibitem [{\citenamefont {Kumar}\ and\ \citenamefont
  {Dasgupta}(2020)}]{Kumar2020}%
  \BibitemOpen
  \bibfield  {author} {\bibinfo {author} {\bibfnamefont {M.}~\bibnamefont
  {Kumar}}\ and\ \bibinfo {author} {\bibfnamefont {C.}~\bibnamefont
  {Dasgupta}},\ }\href
  {https://doi.org/10.1103/PHYSREVE.102.052111/FIGURES/18/MEDIUM} {\bibfield
  {journal} {\bibinfo  {journal} {Phys. Rev. E}\ }\textbf {\bibinfo {volume}
  {102}},\ \bibinfo {pages} {052111} (\bibinfo {year} {2020})}\BibitemShut
  {NoStop}%
\bibitem [{\citenamefont {Agrawal}\ \emph {et~al.}(2022)\citenamefont
  {Agrawal}, \citenamefont {Pandey},\ and\ \citenamefont
  {Mitra}}]{Agrawal2022}%
  \BibitemOpen
  \bibfield  {author} {\bibinfo {author} {\bibfnamefont {V.}~\bibnamefont
  {Agrawal}}, \bibinfo {author} {\bibfnamefont {V.}~\bibnamefont {Pandey}},\
  and\ \bibinfo {author} {\bibfnamefont {D.}~\bibnamefont {Mitra}},\ }\href
  {https://arxiv.org/abs/2206.14172v2} {\bibfield  {journal} {\bibinfo
  {journal} {arXiv:2206.14172}\ } (\bibinfo {year} {2022})}\BibitemShut
  {NoStop}%
\bibitem [{\citenamefont {Anderson}\ \emph {et~al.}(2020)\citenamefont
  {Anderson}, \citenamefont {Glaser},\ and\ \citenamefont
  {Glotzer}}]{Anderson2020}%
  \BibitemOpen
  \bibfield  {author} {\bibinfo {author} {\bibfnamefont {J.~A.}\ \bibnamefont
  {Anderson}}, \bibinfo {author} {\bibfnamefont {J.}~\bibnamefont {Glaser}},\
  and\ \bibinfo {author} {\bibfnamefont {S.~C.}\ \bibnamefont {Glotzer}},\
  }\href {https://doi.org/10.1016/j.commatsci.2019.109363} {\bibfield
  {journal} {\bibinfo  {journal} {Comput. Mater. Sci.}\ }\textbf {\bibinfo
  {volume} {173}},\ \bibinfo {pages} {109363} (\bibinfo {year}
  {2020})}\BibitemShut {NoStop}%
\bibitem [{\citenamefont {Klymko}\ \emph {et~al.}(2018)\citenamefont {Klymko},
  \citenamefont {Geissler}, \citenamefont {Garrahan},\ and\ \citenamefont
  {Whitelam}}]{Klymko2018}%
  \BibitemOpen
  \bibfield  {author} {\bibinfo {author} {\bibfnamefont {K.}~\bibnamefont
  {Klymko}}, \bibinfo {author} {\bibfnamefont {P.~L.}\ \bibnamefont
  {Geissler}}, \bibinfo {author} {\bibfnamefont {J.~P.}\ \bibnamefont
  {Garrahan}},\ and\ \bibinfo {author} {\bibfnamefont {S.}~\bibnamefont
  {Whitelam}},\ }\href {https://doi.org/10.1103/PhysRevE.97.032123} {\bibfield
  {journal} {\bibinfo  {journal} {Phys. Rev. E}\ }\textbf {\bibinfo {volume}
  {97}},\ \bibinfo {pages} {032123} (\bibinfo {year} {2018})}\BibitemShut
  {NoStop}%
\bibitem [{\citenamefont {Ray}\ \emph {et~al.}(2018)\citenamefont {Ray},
  \citenamefont {Chan},\ and\ \citenamefont {Limmer}}]{Ray2018}%
  \BibitemOpen
  \bibfield  {author} {\bibinfo {author} {\bibfnamefont {U.}~\bibnamefont
  {Ray}}, \bibinfo {author} {\bibfnamefont {G.~K.-L.}\ \bibnamefont {Chan}},\
  and\ \bibinfo {author} {\bibfnamefont {D.~T.}\ \bibnamefont {Limmer}},\
  }\href {https://doi.org/10.1063/1.5003151} {\bibfield  {journal} {\bibinfo
  {journal} {J. Chem. Phys.}\ }\textbf {\bibinfo {volume} {148}},\ \bibinfo
  {pages} {124120} (\bibinfo {year} {2018})}\BibitemShut {NoStop}%
\bibitem [{\citenamefont {Das}\ and\ \citenamefont {Limmer}(2019)}]{Das2019a}%
  \BibitemOpen
  \bibfield  {author} {\bibinfo {author} {\bibfnamefont {A.}~\bibnamefont
  {Das}}\ and\ \bibinfo {author} {\bibfnamefont {D.~T.}\ \bibnamefont
  {Limmer}},\ }\href {https://doi.org/10.1063/1.5128956} {\bibfield  {journal}
  {\bibinfo  {journal} {J. Chem. Phys.}\ }\textbf {\bibinfo {volume} {151}},\
  \bibinfo {pages} {244123} (\bibinfo {year} {2019})}\BibitemShut {NoStop}%
\bibitem [{\citenamefont {Whitelam}\ \emph {et~al.}(2020)\citenamefont
  {Whitelam}, \citenamefont {Jacobson},\ and\ \citenamefont
  {Tamblyn}}]{Whitelam2020}%
  \BibitemOpen
  \bibfield  {author} {\bibinfo {author} {\bibfnamefont {S.}~\bibnamefont
  {Whitelam}}, \bibinfo {author} {\bibfnamefont {D.}~\bibnamefont {Jacobson}},\
  and\ \bibinfo {author} {\bibfnamefont {I.}~\bibnamefont {Tamblyn}},\ }\href
  {https://doi.org/10.1063/5.0015301} {\bibfield  {journal} {\bibinfo
  {journal} {J. Chem. Phys.}\ }\textbf {\bibinfo {volume} {153}},\ \bibinfo
  {pages} {044113} (\bibinfo {year} {2020})}\BibitemShut {NoStop}%
\bibitem [{\citenamefont {Ray}\ and\ \citenamefont {Chan}(2020)}]{Ray2020}%
  \BibitemOpen
  \bibfield  {author} {\bibinfo {author} {\bibfnamefont {U.}~\bibnamefont
  {Ray}}\ and\ \bibinfo {author} {\bibfnamefont {G.~K.-L.}\ \bibnamefont
  {Chan}},\ }\href {https://doi.org/10.1063/1.5143144} {\bibfield  {journal}
  {\bibinfo  {journal} {J. Chem. Phys.}\ }\textbf {\bibinfo {volume} {152}},\
  \bibinfo {pages} {104107} (\bibinfo {year} {2020})}\BibitemShut {NoStop}%
\bibitem [{\citenamefont {Helms}\ and\ \citenamefont {Chan}(2020)}]{Helms2020}%
  \BibitemOpen
  \bibfield  {author} {\bibinfo {author} {\bibfnamefont {P.}~\bibnamefont
  {Helms}}\ and\ \bibinfo {author} {\bibfnamefont {G.~K.-L.}\ \bibnamefont
  {Chan}},\ }\href {https://doi.org/10.1103/PhysRevLett.125.140601} {\bibfield
  {journal} {\bibinfo  {journal} {Phys. Rev. Lett.}\ }\textbf {\bibinfo
  {volume} {125}},\ \bibinfo {pages} {140601} (\bibinfo {year}
  {2020})}\BibitemShut {NoStop}%
\bibitem [{\citenamefont {Oakes}\ \emph {et~al.}(2020)\citenamefont {Oakes},
  \citenamefont {Moss},\ and\ \citenamefont {Garrahan}}]{Oakes2020}%
  \BibitemOpen
  \bibfield  {author} {\bibinfo {author} {\bibfnamefont {T.~H.~E.}\
  \bibnamefont {Oakes}}, \bibinfo {author} {\bibfnamefont {A.}~\bibnamefont
  {Moss}},\ and\ \bibinfo {author} {\bibfnamefont {J.~P.}\ \bibnamefont
  {Garrahan}},\ }\href {https://doi.org/10.1088/2632-2153/ab95a1} {\bibfield
  {journal} {\bibinfo  {journal} {Mach. Learn. Sci. Technol}\ }\textbf
  {\bibinfo {volume} {1}},\ \bibinfo {pages} {035004} (\bibinfo {year}
  {2020})}\BibitemShut {NoStop}%
\bibitem [{\citenamefont {Rose}\ \emph {et~al.}(2021)\citenamefont {Rose},
  \citenamefont {Mair},\ and\ \citenamefont {Garrahan}}]{Rose2021}%
  \BibitemOpen
  \bibfield  {author} {\bibinfo {author} {\bibfnamefont {D.~C.}\ \bibnamefont
  {Rose}}, \bibinfo {author} {\bibfnamefont {J.~F.}\ \bibnamefont {Mair}},\
  and\ \bibinfo {author} {\bibfnamefont {J.~P.}\ \bibnamefont {Garrahan}},\
  }\href {https://doi.org/10.1088/1367-2630/abd7bd} {\bibfield  {journal}
  {\bibinfo  {journal} {New J. Phys.}\ }\textbf {\bibinfo {volume} {23}},\
  \bibinfo {pages} {013013} (\bibinfo {year} {2021})}\BibitemShut {NoStop}%
\bibitem [{\citenamefont {Das}\ \emph {et~al.}(2022)\citenamefont {Das},
  \citenamefont {Kuznets-Speck},\ and\ \citenamefont {Limmer}}]{Das2022}%
  \BibitemOpen
  \bibfield  {author} {\bibinfo {author} {\bibfnamefont {A.}~\bibnamefont
  {Das}}, \bibinfo {author} {\bibfnamefont {B.}~\bibnamefont {Kuznets-Speck}},\
  and\ \bibinfo {author} {\bibfnamefont {D.~T.}\ \bibnamefont {Limmer}},\
  }\href {https://journals.aps.org/prl/abstract/10.1103/PhysRevLett.128.028005}
  {\bibfield  {journal} {\bibinfo  {journal} {Phys. Rev. Lett.}\ }\textbf
  {\bibinfo {volume} {128}},\ \bibinfo {pages} {028005} (\bibinfo {year}
  {2022})}\BibitemShut {NoStop}%
\bibitem [{Note1()}]{Note1}%
  \BibitemOpen
  \bibinfo {note} {See Supporting Information, which includes Refs.~\cite
  {Martinez2009, Stukowski2009, Weber2016, Hargus2021, Weeks1971}, for
  additional simulation and calculation details as well as simulation
  movies.}\BibitemShut {Stop}%
\bibitem [{\citenamefont {Schultz}\ and\ \citenamefont
  {Kofke}(2018)}]{Schultz2018}%
  \BibitemOpen
  \bibfield  {author} {\bibinfo {author} {\bibfnamefont {A.~J.}\ \bibnamefont
  {Schultz}}\ and\ \bibinfo {author} {\bibfnamefont {D.~A.}\ \bibnamefont
  {Kofke}},\ }\href {https://doi.org/10.1063/1.5053714} {\bibfield  {journal}
  {\bibinfo  {journal} {J. Chem. Phys}\ }\textbf {\bibinfo {volume} {149}},\
  \bibinfo {pages} {204508} (\bibinfo {year} {2018})}\BibitemShut {NoStop}%
\bibitem [{\citenamefont {Weber}\ \emph {et~al.}(2016)\citenamefont {Weber},
  \citenamefont {Weber},\ and\ \citenamefont {Frey}}]{Weber2016}%
  \BibitemOpen
  \bibfield  {author} {\bibinfo {author} {\bibfnamefont {S.~N.}\ \bibnamefont
  {Weber}}, \bibinfo {author} {\bibfnamefont {C.~A.}\ \bibnamefont {Weber}},\
  and\ \bibinfo {author} {\bibfnamefont {E.}~\bibnamefont {Frey}},\ }\href
  {https://doi.org/https://doi.org/10.1103/PhysRevLett.116.058301} {\bibfield
  {journal} {\bibinfo  {journal} {Phys. Rev. Lett.}\ }\textbf {\bibinfo
  {volume} {116}},\ \bibinfo {pages} {058301} (\bibinfo {year}
  {2016})}\BibitemShut {NoStop}%
\bibitem [{\citenamefont {Hargus}\ \emph {et~al.}(2021)\citenamefont {Hargus},
  \citenamefont {Epstein},\ and\ \citenamefont {Mandadapu}}]{Hargus2021}%
  \BibitemOpen
  \bibfield  {author} {\bibinfo {author} {\bibfnamefont {C.}~\bibnamefont
  {Hargus}}, \bibinfo {author} {\bibfnamefont {J.~M.}\ \bibnamefont
  {Epstein}},\ and\ \bibinfo {author} {\bibfnamefont {K.~K.}\ \bibnamefont
  {Mandadapu}},\ }\href
  {https://journals.aps.org/prl/abstract/10.1103/PhysRevLett.127.178001}
  {\bibfield  {journal} {\bibinfo  {journal} {Phys. Rev. Lett.}\ }\textbf
  {\bibinfo {volume} {127}},\ \bibinfo {pages} {178001} (\bibinfo {year}
  {2021})}\BibitemShut {NoStop}%
\bibitem [{Note2()}]{Note2}%
  \BibitemOpen
  \bibinfo {note} {The low overall density results in these dense liquids
  occupying little volume, forming spherical droplets within the dilute fluid
  phase, which comprises most of the system by volume.}\BibitemShut {Stop}%
\bibitem [{\citenamefont {Aifantis}\ and\ \citenamefont
  {Serrin}(1983)}]{Aifantis1983a}%
  \BibitemOpen
  \bibfield  {author} {\bibinfo {author} {\bibfnamefont {E.~C.}\ \bibnamefont
  {Aifantis}}\ and\ \bibinfo {author} {\bibfnamefont {J.~B.}\ \bibnamefont
  {Serrin}},\ }\href {https://doi.org/10.1016/0021-9797(83)90054-1} {\bibfield
  {journal} {\bibinfo  {journal} {J. Colloid Interface Sci.}\ }\textbf {\bibinfo
  {volume} {96}},\ \bibinfo {pages} {530} (\bibinfo {year} {1983})}\BibitemShut
  {NoStop}%
\bibitem [{\citenamefont {Steinhardt}\ \emph {et~al.}(1983)\citenamefont
  {Steinhardt}, \citenamefont {Nelson},\ and\ \citenamefont
  {Ronchetti}}]{Steinhardt1983}%
  \BibitemOpen
  \bibfield  {author} {\bibinfo {author} {\bibfnamefont {P.~J.}\ \bibnamefont
  {Steinhardt}}, \bibinfo {author} {\bibfnamefont {D.~R.}\ \bibnamefont
  {Nelson}},\ and\ \bibinfo {author} {\bibfnamefont {M.}~\bibnamefont
  {Ronchetti}},\ }\href {https://doi.org/10.1103/PhysRevB.28.784} {\bibfield
  {journal} {\bibinfo  {journal} {Phys. Rev. B}\ }\textbf {\bibinfo {volume}
  {28}},\ \bibinfo {pages} {784} (\bibinfo {year} {1983})}\BibitemShut
  {NoStop}%
\bibitem [{\citenamefont {Omar}\ \emph {et~al.}(2021)\citenamefont {Omar},
  \citenamefont {Klymko}, \citenamefont {GrandPre},\ and\ \citenamefont
  {Geissler}}]{Omar2021}%
  \BibitemOpen
  \bibfield  {author} {\bibinfo {author} {\bibfnamefont {A.~K.}\ \bibnamefont
  {Omar}}, \bibinfo {author} {\bibfnamefont {K.}~\bibnamefont {Klymko}},
  \bibinfo {author} {\bibfnamefont {T.}~\bibnamefont {GrandPre}},\ and\
  \bibinfo {author} {\bibfnamefont {P.~L.}\ \bibnamefont {Geissler}},\ }\href
  {https://doi.org/10.1103/PhysRevLett.126.188002} {\bibfield  {journal}
  {\bibinfo  {journal} {Phys. Rev. Lett.}\ }\textbf {\bibinfo {volume} {126}},\
  \bibinfo {pages} {188002} (\bibinfo {year} {2021})}\BibitemShut {NoStop}%
\bibitem [{Note3()}]{Note3}%
  \BibitemOpen
  \bibinfo {note} {We note that our simulation duration ($5000\tau $) limited
  the periods that we could observe to those with $\tau _{\protect \rm ts} <
  2500\tau $, which coincides with $\Delta \geq 1.2$.}\BibitemShut {Stop}%
\bibitem [{\citenamefont {Stenhammar}\ \emph {et~al.}(2015)\citenamefont
  {Stenhammar}, \citenamefont {Wittkowski}, \citenamefont {Marenduzzo},\ and\
  \citenamefont {Cates}}]{Stenhammar2015}%
  \BibitemOpen
  \bibfield  {author} {\bibinfo {author} {\bibfnamefont {J.}~\bibnamefont
  {Stenhammar}}, \bibinfo {author} {\bibfnamefont {R.}~\bibnamefont
  {Wittkowski}}, \bibinfo {author} {\bibfnamefont {D.}~\bibnamefont
  {Marenduzzo}},\ and\ \bibinfo {author} {\bibfnamefont {M.~E.}\ \bibnamefont
  {Cates}},\ }\href {https://doi.org/10.1103/PhysRevLett.114.018301} {\bibfield
   {journal} {\bibinfo  {journal} {Phys. Rev. Lett.}\ }\textbf {\bibinfo
  {volume} {114}},\ \bibinfo {pages} {018301} (\bibinfo {year}
  {2015})}\BibitemShut {NoStop}%
\bibitem [{\citenamefont {Wysocki}\ \emph {et~al.}(2016)\citenamefont
  {Wysocki}, \citenamefont {Winkler},\ and\ \citenamefont
  {Gompper}}]{Wysocki2016}%
  \BibitemOpen
  \bibfield  {author} {\bibinfo {author} {\bibfnamefont {A.}~\bibnamefont
  {Wysocki}}, \bibinfo {author} {\bibfnamefont {R.~G.}\ \bibnamefont
  {Winkler}},\ and\ \bibinfo {author} {\bibfnamefont {G.}~\bibnamefont
  {Gompper}},\ }\href {https://doi.org/10.1088/1367-2630/AA529D} {\bibfield
  {journal} {\bibinfo  {journal} {New J. Phys.}\ }\textbf {\bibinfo {volume}
  {18}},\ \bibinfo {pages} {123030} (\bibinfo {year} {2016})}\BibitemShut
  {NoStop}%
\bibitem [{\citenamefont {Wittkowski}\ \emph {et~al.}(2017)\citenamefont
  {Wittkowski}, \citenamefont {Stenhammar},\ and\ \citenamefont
  {Cates}}]{Wittkowski2017}%
  \BibitemOpen
  \bibfield  {author} {\bibinfo {author} {\bibfnamefont {R.}~\bibnamefont
  {Wittkowski}}, \bibinfo {author} {\bibfnamefont {J.}~\bibnamefont
  {Stenhammar}},\ and\ \bibinfo {author} {\bibfnamefont {M.~E.}\ \bibnamefont
  {Cates}},\ }\href {https://doi.org/10.1088/1367-2630/aa8195} {\bibfield
  {journal} {\bibinfo  {journal} {New J. Phys.}\ }\textbf {\bibinfo {volume}
  {19}},\ \bibinfo {pages} {105003} (\bibinfo {year} {2017})}\BibitemShut
  {NoStop}%
\bibitem [{\citenamefont {Omar}\ \emph {et~al.}(2019)\citenamefont {Omar},
  \citenamefont {Wu}, \citenamefont {Wang},\ and\ \citenamefont
  {Brady}}]{Omar2019}%
  \BibitemOpen
  \bibfield  {author} {\bibinfo {author} {\bibfnamefont {A.~K.}\ \bibnamefont
  {Omar}}, \bibinfo {author} {\bibfnamefont {Y.}~\bibnamefont {Wu}}, \bibinfo
  {author} {\bibfnamefont {Z.-G.}\ \bibnamefont {Wang}},\ and\ \bibinfo
  {author} {\bibfnamefont {J.~F.}\ \bibnamefont {Brady}},\ }\href
  {https://doi.org/10.1021/acsnano.8b07421} {\bibfield  {journal} {\bibinfo
  {journal} {ACS Nano}\ }\textbf {\bibinfo {volume} {13}},\ \bibinfo {pages}
  {560} (\bibinfo {year} {2019})}\BibitemShut {NoStop}%
\bibitem [{Note4()}]{Note4}%
  \BibitemOpen
  \bibinfo {note} {The diagonal elements of the mutual diffusion tensor are not
  to be confused with the self diffusion coefficients computed in this work.
  The mutual diffusion tensor describes the system response to imposed
  gradients, and its elements are only equal to self-diffusion coefficients in
  the dilute limit.}\BibitemShut {Stop}%
\bibitem [{\citenamefont {Mart{\'{i}}nez}\ \emph {et~al.}(2009)\citenamefont
  {Mart{\'{i}}nez}, \citenamefont {Andrade}, \citenamefont {Birgin},\ and\
  \citenamefont {Mart{\'{i}}nez}}]{Martinez2009}%
  \BibitemOpen
  \bibfield  {author} {\bibinfo {author} {\bibfnamefont {L.}~\bibnamefont
  {Mart{\'{i}}nez}}, \bibinfo {author} {\bibfnamefont {R.}~\bibnamefont
  {Andrade}}, \bibinfo {author} {\bibfnamefont {E.~G.}\ \bibnamefont
  {Birgin}},\ and\ \bibinfo {author} {\bibfnamefont {J.~M.}\ \bibnamefont
  {Mart{\'{i}}nez}},\ }\href {https://doi.org/10.1002/jcc.21224} {\bibfield
  {journal} {\bibinfo  {journal} {J. Comput. Chem.}\ }\textbf {\bibinfo
  {volume} {30}},\ \bibinfo {pages} {2157} (\bibinfo {year}
  {2009})}\BibitemShut {NoStop}%
\bibitem [{\citenamefont {Stukowski}(2009)}]{Stukowski2009}%
  \BibitemOpen
  \bibfield  {author} {\bibinfo {author} {\bibfnamefont {A.}~\bibnamefont
  {Stukowski}},\ }\href {https://doi.org/10.1088/0965-0393/18/1/015012}
  {\bibfield  {journal} {\bibinfo  {journal} {Model. Numer. Simul. Mater. Sci.}\
  }\textbf {\bibinfo {volume} {18}},\ \bibinfo {pages} {015012} (\bibinfo
  {year} {2009})}\BibitemShut {NoStop}%
\bibitem [{\citenamefont {Weeks}\ \emph {et~al.}(1971)\citenamefont {Weeks},
  \citenamefont {Chandler},\ and\ \citenamefont {Andersen}}]{Weeks1971}%
  \BibitemOpen
  \bibfield  {author} {\bibinfo {author} {\bibfnamefont {J.~D.}\ \bibnamefont
  {Weeks}}, \bibinfo {author} {\bibfnamefont {D.}~\bibnamefont {Chandler}},\
  and\ \bibinfo {author} {\bibfnamefont {H.~C.}\ \bibnamefont {Andersen}},\
  }\href {https://doi.org/10.1063/1.1674820} {\bibfield  {journal} {\bibinfo
  {journal} {J. Chem. Phys.}\ }\textbf {\bibinfo {volume} {54}},\ \bibinfo
  {pages} {5237} (\bibinfo {year} {1971})}\BibitemShut {NoStop}%
\end{thebibliography}
\end{document}


\title{Supporting Information -- Phase Coexistence Implications of Violating Newton's Third Law}

\author{Yu-Jen Chiu}
\affiliation{Department of Materials Science and Engineering, University of California, Berkeley, California 94720, USA}

\author{Ahmad K. Omar}
\email{aomar@berkeley.edu}
\affiliation{Department of Materials Science and Engineering, University of California, Berkeley, California 94720, USA}
\affiliation{Materials Sciences Division, Lawrence Berkeley National Laboratory, Berkeley, California 94720, USA}

\maketitle
\section{Supporting Videos}
The videos included in the Supporting Information are intended to serve as representative examples of the states/transitions listed below. 
For reference to the expected globally stable states as a function of ($\Delta$, $\phi$), see the phase diagram provided in the main text [Fig.~1(b)]. 
In all videos, the captioned time is in units of $\tau$. 
All videos are freely available at:\newline
\url{https://berkeley.box.com/s/ne2deyear7sand2pm5ybht16yryuwd98}
\begin{enumerate}
    \item \textbf{Three-fluid Coexistence Coarsening Dynamics} \newline three\_fluid\_coexistence\_formation.mp4 \newline ($\Delta=0.3$, $\phi=0.0875$) \newline Illustrates the time evolution of an initially homogeneous fluid to a state of three-fluid coexistence.
    \item \textbf{Strongly Segregated Liquid-gas Coexistence Coarsening Dynamics} \newline demix\_liquid\_gas\_formation.mp4 \newline ($\Delta=0.6$, $\phi=0.1$) \newline Illustrates the time evolution of an initially homogeneous fluid to a state of liquid-gas coexistence with strong species segregation.
    \item \textbf{Homogeneous Clustering Stationary Dynamics} \newline homogeneous\_clustering\_dynamic.mp4 \newline ($\Delta=0.5$, $\phi=0.2$) \newline Illustrates the stationary dynamics of the homogeneous clustering state. 
    \item \textbf{Traveling Liquid} \newline traveling\_liquid-gas\_coexistence.mp4 \newline ($\Delta=0.35$, $\phi=0.5$) \newline Illustrates the oscillatory dynamics of the traveling liquid state.
    \item \textbf{Traveling Crystal at Low Nonreciprocity} \newline traveling\_crystal-gas\_low\_nonreciprocity.mp4 \newline ($\Delta=0.9$, $\phi=0.5$) \newline Illustrates the dynamics of the traveling crystal at low nonreciprocity. At low nonreciprocity, there is no discernible oscillatory behavior of the center-of-mass dynamics.
    \item \textbf{Traveling Crystal at High Nonreciprocity} \newline traveling\_crystal-gas\_high\_nonreciprocity.mp4 \newline ($\Delta=1.5$, $\phi=0.5$) \newline Illustrates the dynamics of the traveling crystal at high nonreciprocity. At high nonreciprocity, several periods of oscillation are observed within the simulation duration.
\end{enumerate}

\section{Assessing Global Stability}
The phase diagram reported in the main text reports the stable configuration from simulations initialized with a random configuration generated with Packmol~\cite{Martinez2009} and run for a duration of $5000\tau$.
To determine whether the observed final state of the system was indeed representative of the globally stable system configuration, we prepared systems with differing initial configurations (detailed below) to observe if the same final configuration is reached, independent of the initial configuration. 
Reaching the same final configuration from disparate initial configurations is a necessary requirement for establishing global stability that, in the absence of robust tools for the importance sampling of many-body nonequilibrium systems, is also the most direct method for exploring the stability of driven systems. 

The primary areas of our phase diagram where the globally stable state is ambiguous are, much like in equilibrium systems, near the reported boundaries between differing states in our phase diagram. 
Near phase boundaries, particularly for multicomponent systems, systems are often replete with a number of metastable states with significant barriers precluding the facile exploration of these states.  
Here, we discuss our findings near the boundary between three-phase and two-phase coexistence as the boundaries between these states proved difficult to establish. 
We focus on a global volume fraction $\phi = 0.1$. 
In addition to a random and homogeneous initial configuration, we utilize an initial configuration of a liquid (i.e.,~the particle configuration within the liquid is amorphous) spherical droplet of pure $L$ consisting of $5000$ $L$ particles (corresponding to $10\%$ of the total system particles) with a local volume fraction of $\phi_{\rm sphere} = 0.51$ and a droplet radius of $12\sigma$.

For $\Delta = 0.25$ to $0.25\leq\Delta\leq0.4$, seeding our system with a large $L$ droplet was found to result in three fluid phase coexistence rather than the two-fluid coexistence observed when beginning from a homogeneous initial configuration. 
The two-fluid coexistence observed beginning from the original configuration consists of a $G$-rich dense liquid phase. 
While the dilute phase is enriched in $L$, nucleating an $L$ droplet from the dilute phase and achieving a state of three-fluid coexistence may simply be too unlikely given the timescale of our simulations.
We further note that for $0.5\leq\Delta\leq1.5$, the initialized $L$ droplet is indeed found to dissolve, but the system remains homogeneous and appears more similar to the homogeneous clustering regime than a state of two-fluid coexistence. 
 
Seeding with a spherical droplet of pure $G$, consist of $5000$ $G$ particles with $\phi_{\rm sphere} = 0.51$ and a droplet radius of $12\sigma$ also, is found to result in final configurations that are more similar to those reached with random initial configurations, with three-fluid coexistence observed for $0.2\leq\Delta\leq0.3$ and states of strongly segregated two-fluid coexistence for $0.5\leq\Delta\leq1.5$ (eliminating the high $\Delta$ homogeneous clustering state).
Taken together, these results emphasize that the boundaries between phases, where global stability cannot be precisely established, cannot be interpreted strictly.

\section{Two-Phase Coexistence Characterization}

\begin{figure}
	\centering
	\includegraphics[width=.70\textwidth]{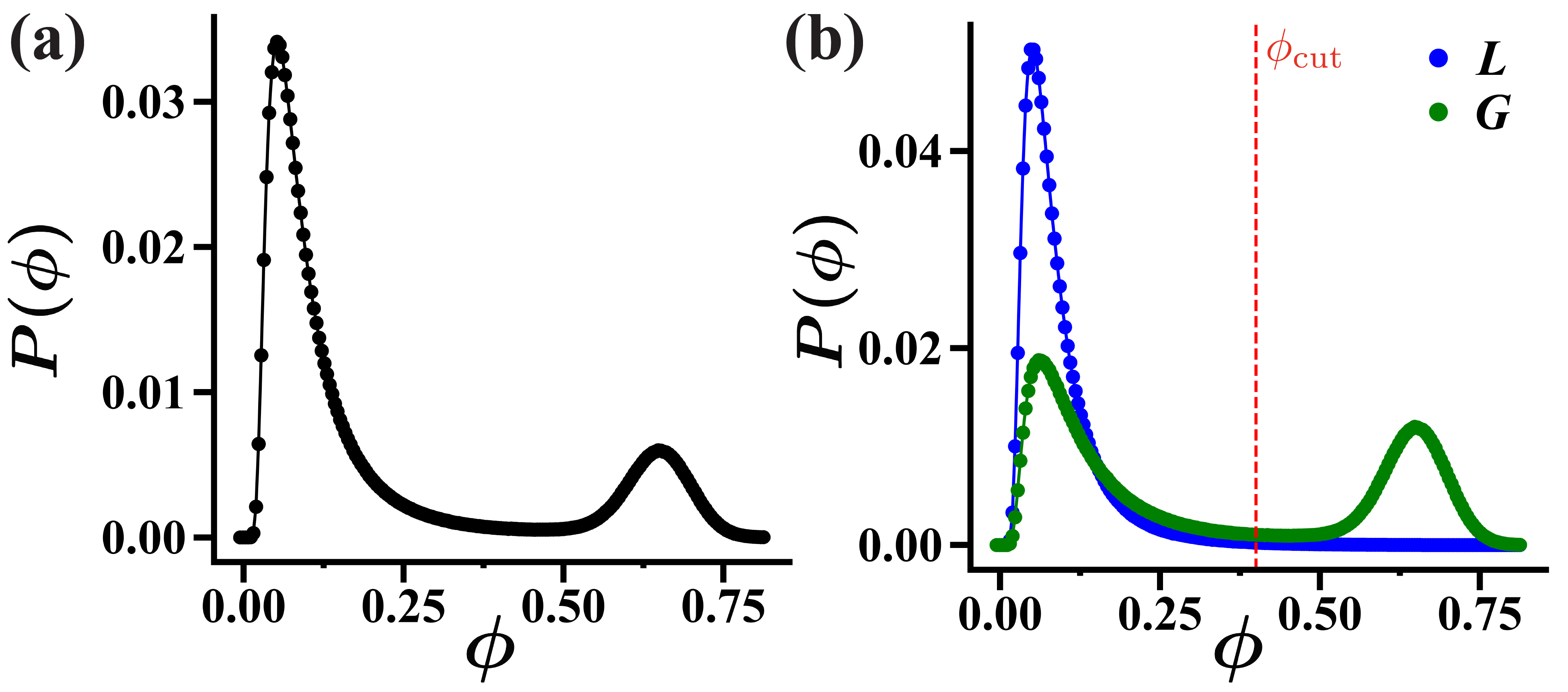}
	\caption{\protect\small{{Example per-particle volume fraction probabilities for a state of two-phase coexistence with a global density of $\phi = 0.0875$  and $\Delta = 1.1$. The distribution computed over particles of (a) all species and (b) each species separately. Here, $\phi_{\rm cut} = 0.4$. As our mixture is symmetric ($\chi = 0.5$), panel (a) is simply the average of the curves presented in panel (b).}}}
	\label{sfig:densitydistribution}
\end{figure}
The coexistence densities and compositions presented in the main text [Fig.~2] are obtained by examining the distributions of the local per-particle volume fraction [see Fig.~\ref{sfig:densitydistribution}].
This per-particle volume fraction is found by computing the local Voronoi volume of each particle $V_{\text{Vor}}$~\cite{Stukowski2009}.
The local volume fraction of an individual particle is then simply $\phi = \pi d_{\rm rec}^3/6V_{\text{Vor}}$. 
In regions of liquid-gas phase separation, the distribution is bimodal [see Fig.~\ref{sfig:densitydistribution}(a)], and the gas and liquid densities are taken to be the densities at the two local maxima.
The compositions of the two phases are then obtained by considering the local density distribution of each species separately [see Fig.~\ref{sfig:densitydistribution}(b)]. 
All particles above (below) a cutoff density, $\phi_{\rm cut}$, that is between the two coexisting densities are assigned to the liquid (gas) phase.
In this way, we can determine the number of particles of each species in each phase and therefore determine each phase's composition. 

\section{Mapping Low-Nonreciprocity States to Passive Systems with Effective Interactions or Temperatures}

\begin{figure}
	\centering
	\includegraphics[width=.7\textwidth]{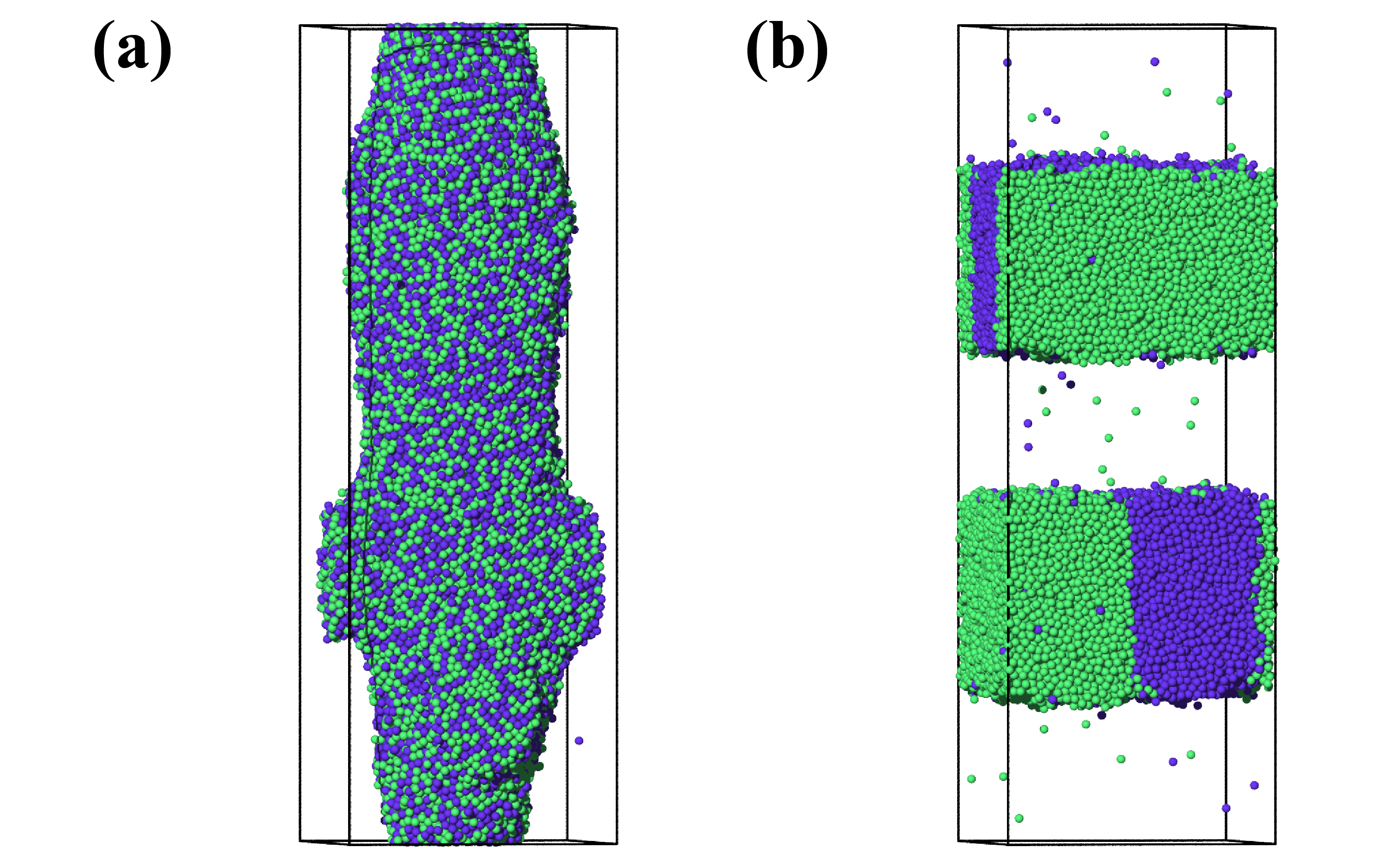}
	\caption{\protect\small{{Simulation snapshots of final configurations reached upon of alteration of the mutual interaction energy, (a) $\varepsilon_{\rm LG}^{\rm eff} > \varepsilon$ and (b) $\varepsilon_{\rm LG}^{\rm eff} < \varepsilon$ at $\phi = 0.3$, $\chi = 0.5$ and $\varepsilon = 2.0$. Note that the regions would continue to reduce the inerfacial area separating the phases with additional simulation time. }}}
	\label{sfig:interactenergysnap}
\end{figure}

One way to perhaps understand the low-reciprocity states of coexistence state is by mapping the system to an equilibrium analogue with modified interaction potentials. 
As discussed in the main text, there is no clear reason to modify the effective interactions between like-species as nonreciprocity only affects forces between dissimilar species. 
The effective interspecies interaction energy, $\varepsilon_{LG}$, must therefore contain the effects of nonreciprocity should a mapping hold.  
Figure~\ref{sfig:interactenergysnap} illustrates the effects of altering $\varepsilon_{LG}$ for a passive system with fixed intraspecies interaction energies $\varepsilon_{LL} = \varepsilon_{GG} = \varepsilon = 2.0$ beginning from a random initial configuration.
Increasing the magnitude of the interspecies interaction energy $\varepsilon^{\rm eff}_{LG} > \varepsilon$ [see Fig.~\ref{sfig:interactenergysnap}(a)] results in a state of liquid-gas coexistence with identical compositions $\chi_{\rm gas} =\chi_{\rm liq} = 0.5$.
As anticipated, the composition between two coexisting phases cannot be altered by simply changing $\varepsilon_{LG}$. 
However, unlike states of two-phase coexistence, which require differences in like-species interaction energies (i.e.,~$\varepsilon_{LL} \neq \varepsilon_{GG}$)
to generate composition differences between phases in a symmetric mixture ($\chi = 0.5$), altering the \textit{interspecies} interaction energy is all that is required to generate a state of three-fluid coexistence with strong species segregation. 
Lowering the interaction energy magnitude $\varepsilon^{\rm eff}_{LG} < \varepsilon$ can generate three-phase coexistence with sharply contrasting compositions.
The three phases consist of two dense liquids nearly pure in either species $L$ or $G$, and a less dense phase with equal amount of each species.
Thus, while two-phase coexistence scenarios cannot be mapped to equilibrium systems, there is a possibility that our observed nonequilibrium three-fluid coexistence can be mapped to an equilibrium system.
A derivation linking the modified interspecies interaction energy to nonreciprocity is left to future work. 

\begin{figure}
	\centering
	\includegraphics[width=.35\textwidth]{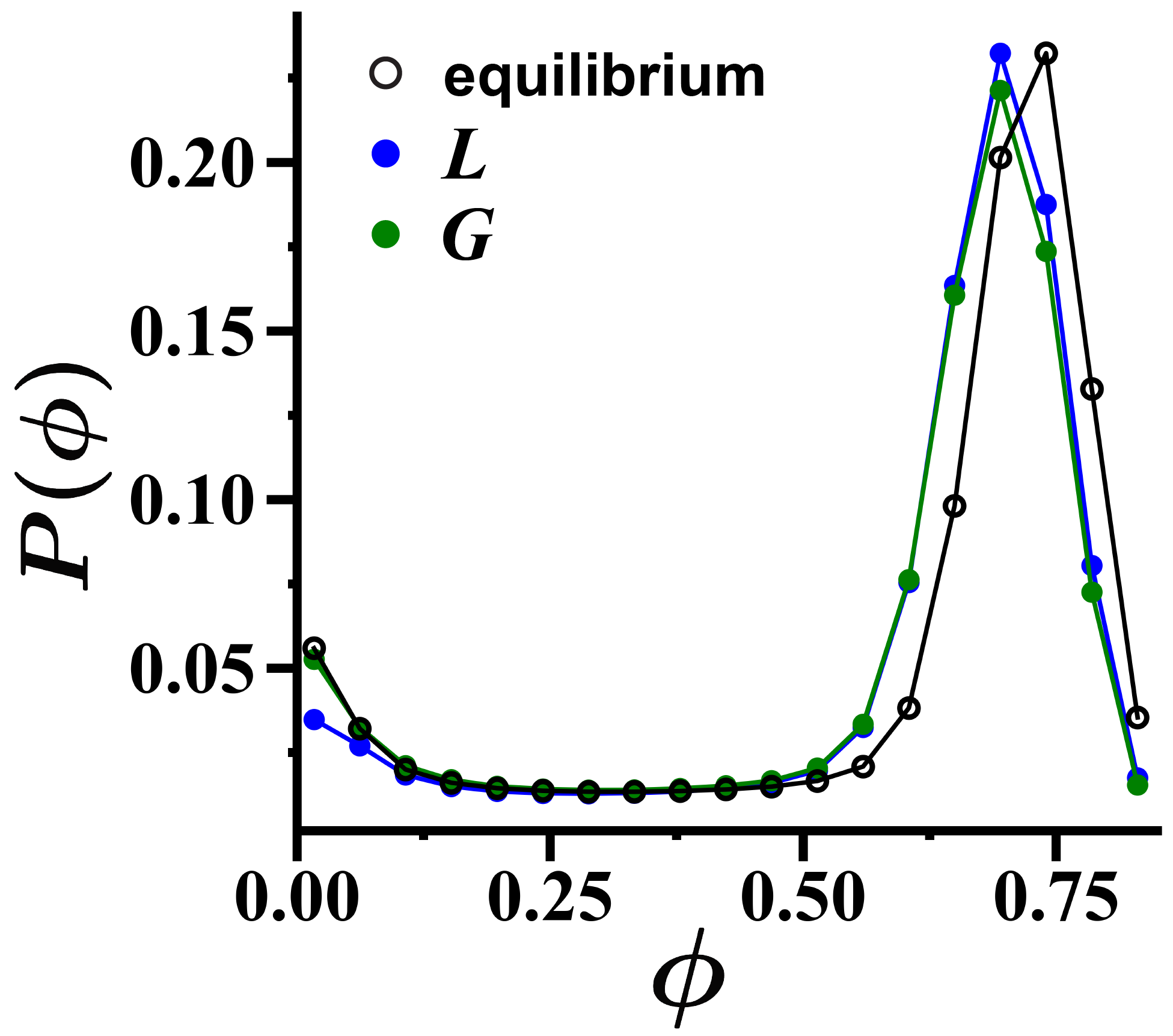}
	\caption{\protect\small{{Per-particle volume fraction probability for the equilibrium system ($\Delta = 0$) and the effective species temperature system (with $T_G/T = 1.2$ and $T_L/T = 1.1$) with a global density and composition of $\phi = 0.0875$ and $\chi = 0.5$.}}}
	\label{sfig:effectivetemp}
\end{figure}

While two-phase coexistence at low nonreciprocity is unlikely to be described by a passive system with modified interactions, perhaps it is possible to make a simpler nonequilibrium system analogy. 
For instance, it is instructive to consider if the influence of nonreciprocity is to alter the effective temperatures experienced by particles $G$ and $L$.
Indeed, particles of different species are found to have different self diffusion constants, with particles of species $G$ experiencing a greater diffusivity than particles of species $L$.
An appeal to the Stokes-Einstein relation would lead to the conclusion that, in this mapping, $T_G > T_L$. 
However, simulating a system in which all interactions are all reciprocal (and taken to be temperature independent) but with particles of different species  experiencing different temperatures is found to drive the species at higher temperature ($G$) into the gas phase (see Fig.~\ref{sfig:effectivetemp}), aligning with the result reported in previous work with demix of particles with different diffusivity~\cite{Weber2016}. 
At low nonreciprocity, we observe the opposite trend: the gas phase is enriched in $L$ particles, suggesting that the effective temperature mapping is a poor description of the system.
A true nonequilibrium multicomponent coexistence theory will be required to understand the observed phase behavior.

\section{Traveling State Characterization}
\begin{figure}
	\centering
	\includegraphics[width=.80\textwidth]{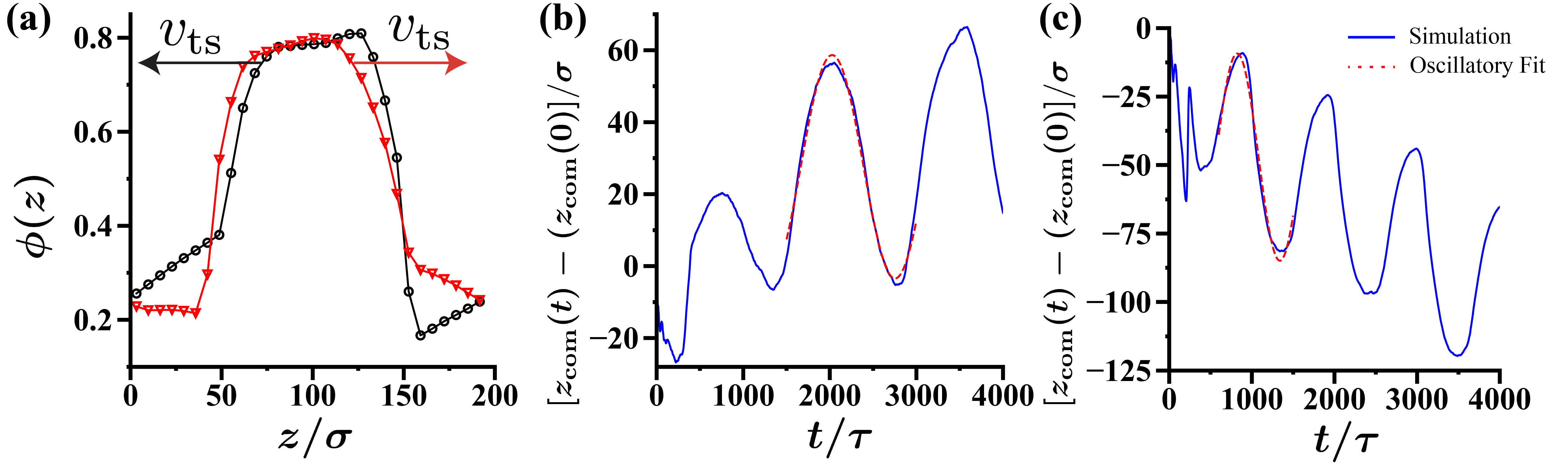}
	\caption{\protect\small{{Additional characterization of the traveling crystal-fluid coexistence states with $\phi = 0.5$. (a) Two  instantaneous (averaging time $< \tau_{\rm ts}$) density profiles with opposite traveling directions for traveling crystal-fluid coexistence states for $\Delta = 1.5$. Relative center-of-mass position and the fit to an oscillatory center-of-mass velocity for (b) $\Delta = 1.2$ and (c) $\Delta = 1.5$.}}}
	\label{sfig:COMposition}
\end{figure}

While a long-time average ($> \tau_{\rm ts}$) of the traveling states' density profile returns a stationary-like profile (see Fig.~4(a) in the main text), the instantaneous density profiles are markedly different.
Figure~\ref{sfig:COMposition}(a) shows representative density profiles at two well-separated points in time.
While the crystal phase is relatively homogeneous, the dilute fluid exhibits a nearly constant density gradient.

The relative center-of-mass positions are tracked as a function of time, and are used to fit the traveling center-of-mass velocity $v_{\rm com}$ to an oscillatory form [see Figs.~\ref{sfig:COMposition}(b) and (c)].
We note that the displacements are not purely oscillatory as there remains a diffusive element to the dynamics that results in a net displacement over long times.
This diffusive drift is neglected in fitting the oscillatory form of the velocity. 

\section{Characterization of Homogeneous States}
The theoretical description of the phase behavior of any material will require an understanding of the homogeneous state (i.e.,~states where long wavelength perturbations to the spatial density or composition profiles are stable).
While phase separated and traveling states make up much of our phase diagram, there is a broad continuous region in the $\Delta-\phi$ plane where long wavelength fluctuations are stable which we term the ``homogeneous clustering regime."
Understanding and characterizing this regime will be essential towards our understanding of nonreciprocal phase transitions. 
To this end, here we characterize the structure and dynamics of these states which we anticipate will aid in the theoretical development of multicomponent nonequilibrium phase behavior. 

\subsection{Structural Features}

\begin{figure}
	\centering
	\includegraphics[width=.35\textwidth]{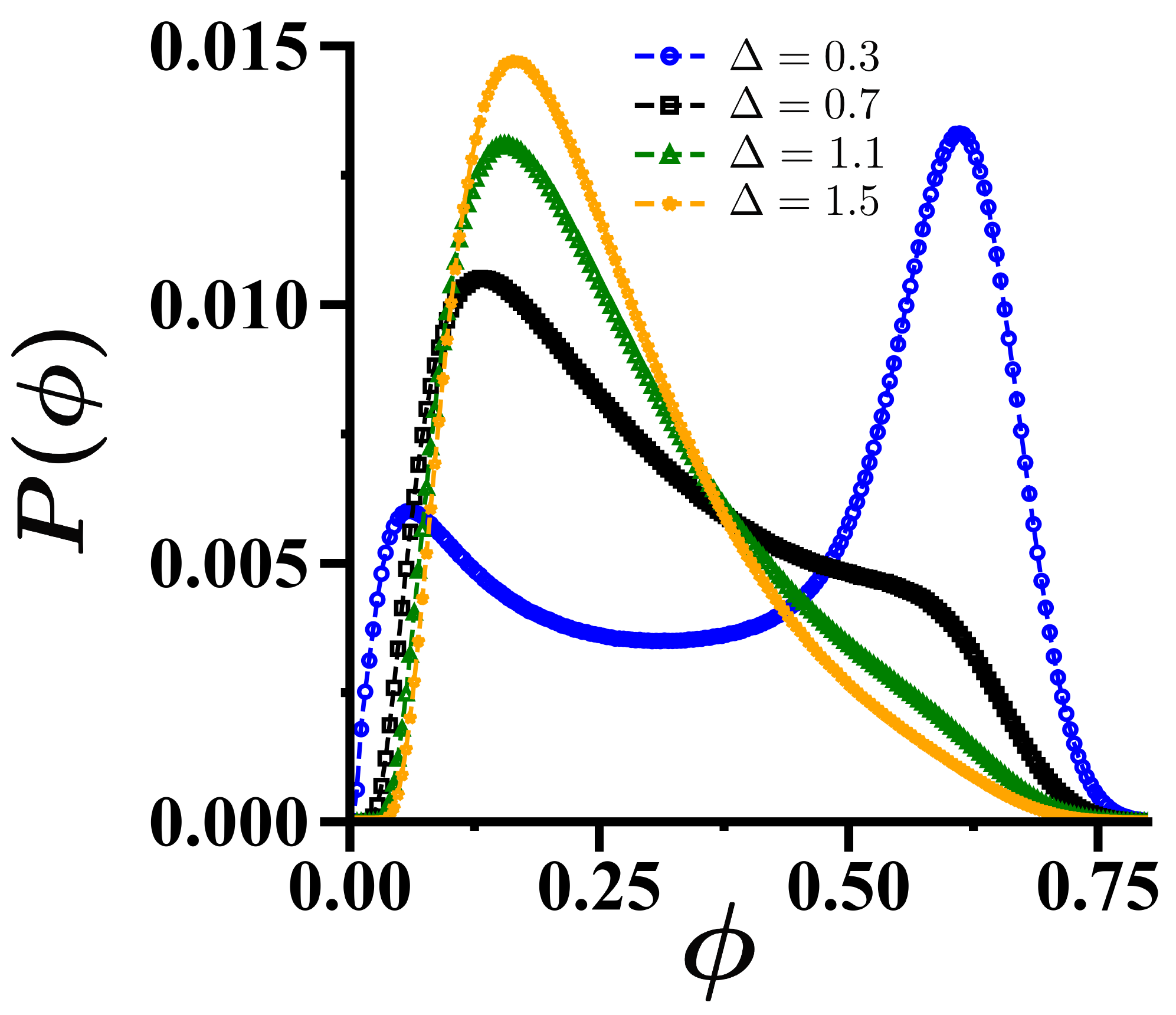}
	\caption{\protect\small{{Nonreciprocity dependence of the per-particle volume fraction probability distributions with global density $\phi = 0.2$. }}}
	\label{sfig:localvolume}
\end{figure}

We begin by examining some coarse-measures for the structural changes brought on by nonreciprocity. 
It is instructive to examine the full distribution of the local particle volume fraction (obtained by using the per-particle Vornonoi volumes) in the homogeneous regime.
For $\phi = 0.2$, at the lowest nonreciprocity, the system is approaching the phase separated regime (see Fig.~1(b) in the main text) and, consequently, the density distribution appears to be bimodal, as shown in Figure~\ref{sfig:localvolume}.
With increasing $\Delta$, the high density peak is eliminated, with the low density peak remaining with a broad shoulder. 
It is interesting to note that despite the system appearing homogeneous with long-range correlations all but extinguished at high $\Delta$, the local density distribution remains decidedly non-Gaussian.
The origins of this distribution will be explored further in future work.

\begin{figure}
	\centering
	\includegraphics[width=.7\textwidth]{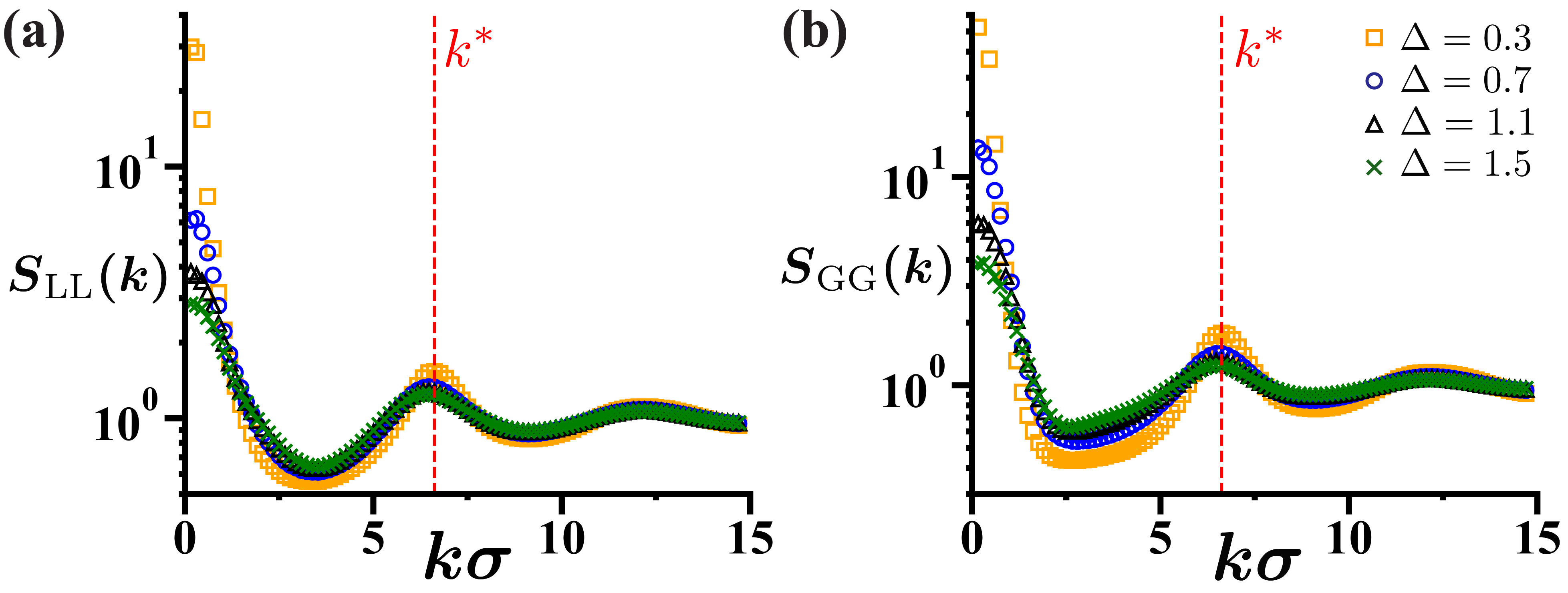}
	\caption{\protect\small{{Nonreciprocity dependence of the partial static structure factors for (a) $LL$ and (b) $GG$ density correlations with $\phi = 0.2$. All curves exhibit a peak located at $k^*$, which corresponds to a length scale of the reciprocity diameter (i.e.,~$k^* \approx 2\pi/d_{\rm rec}$).}}}
	\label{sfig:structurefactor}
\end{figure}

We now compute the partial static structure factors $S_{\alpha\beta}$ to understand the spatial correlations between the Fourier transformed densities of species $\alpha$ and $\beta$. 
Here, we focus on \textit{intraspecies} densities correlations (i.e.,~$\alpha=\beta$) with $S_{\alpha\alpha}$ taking the same form presented in the main text method section.
We present results for $S_{LL}(k)$ and $S_{GG}(k)$ (where $k = |\mathbf{k}|$) in Fig.~\ref{sfig:structurefactor}, which illustrates that long wavelength (i.e.,~$k\rightarrow 0$) density fluctuations are increasingly suppressed with increasing nonreciprocity. 

\begin{figure}
	\centering
	\includegraphics[width=.35\textwidth]{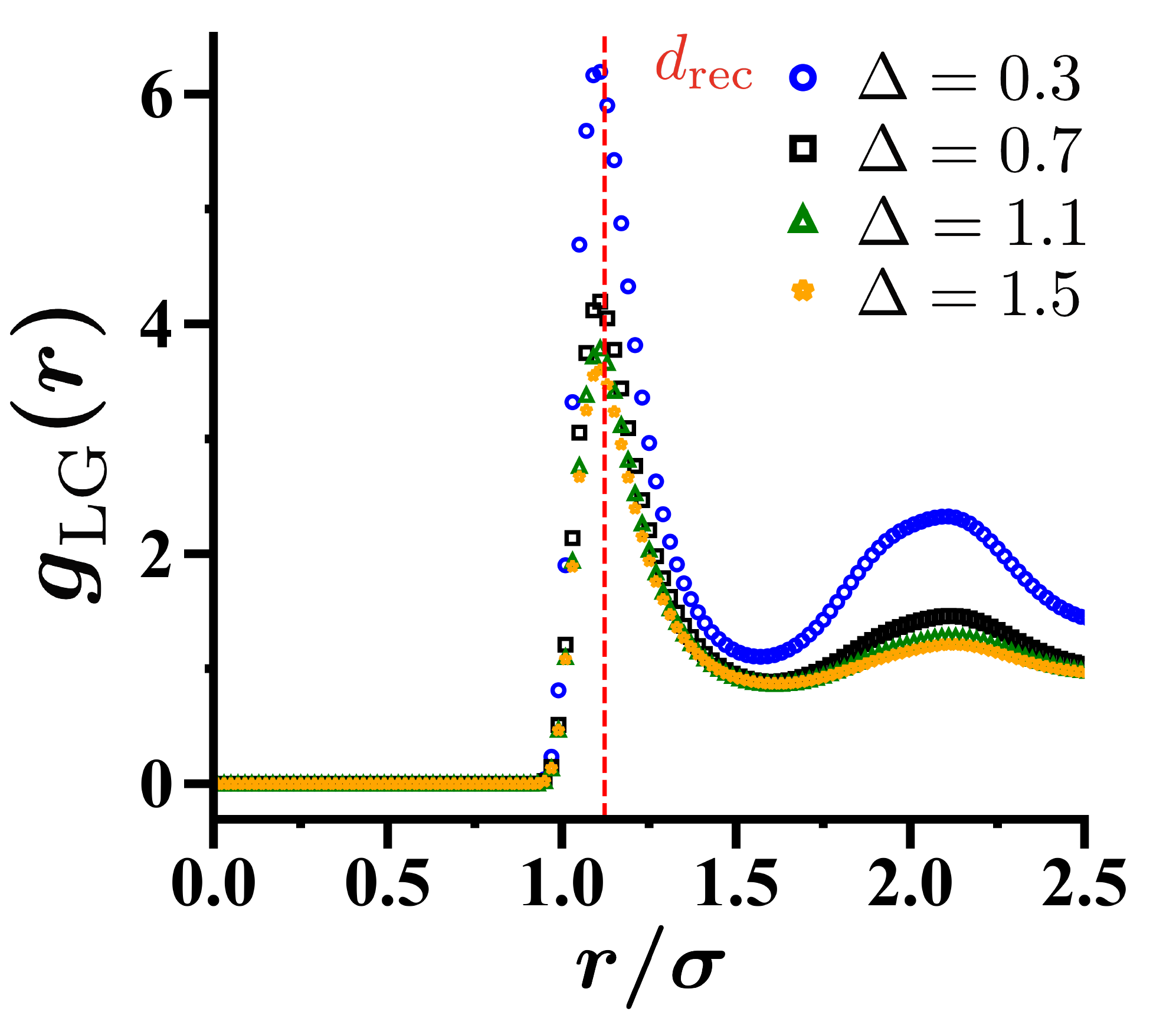}
	\caption{\protect\small{{Nonreciprocity dependence of the interspecies radial distribution with global density $\phi = 0.2$. }}}
	\label{sfig:rdf}
\end{figure}

It is more convenient to examine the changes in short-ranged structural changes through the radial distribution function (RDF). 
The \textit{interspecies} RDF between species $L$ and $G$ is shown in Fig.~\ref{sfig:rdf}. 
With increasing $\Delta$ increase, short-ranged correlations between $L$ and $G$ particles are diminished. 
The dissolution of large clusters not only reduced long wavelength correlations, but short-ranged correlations were also necessarily reduced.  

\subsection{Dynamical Features}
\begin{figure}
	\centering
	\includegraphics[width=.35\textwidth]{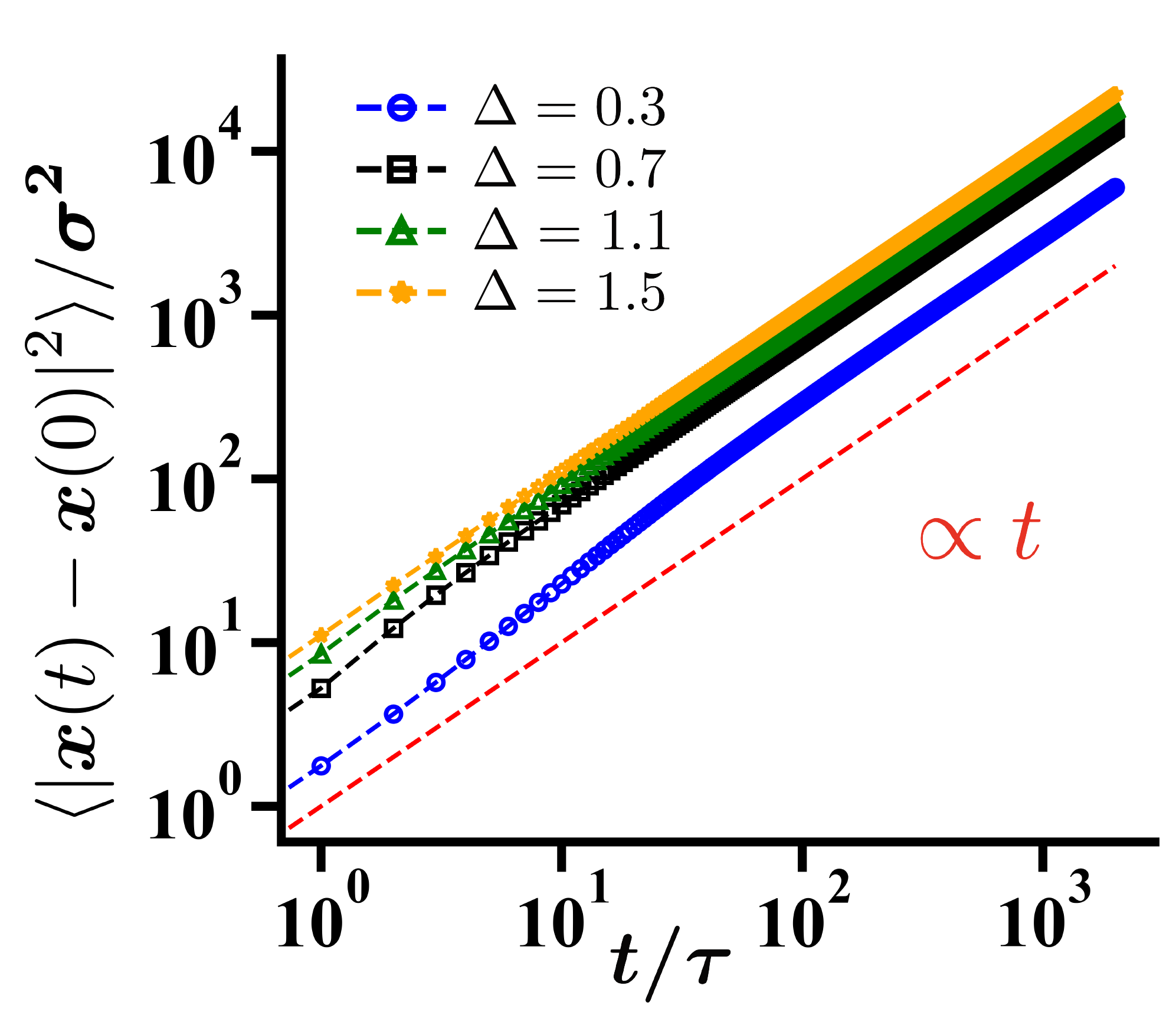}
	\caption{\protect\small{{Nonreciprocity dependence of the mean-square displacement with global density $\phi = 0.2$.}}}
	\label{sfig:msd}
\end{figure}

To characterize the dynamics of the homogeneous state, we compute the self diffusion coefficient of each species from the mean-square displacement (MSD):
\begin{equation}
    \label{msddiffusion}
    D^{\rm self}_{\alpha} = \frac{1}{2d}\lim_{t\rightarrow\infty}\frac{d}{dt}\langle |\mathbf{x}(t)-\mathbf{x}(0)|^2 \rangle_\alpha
\end{equation}
where $\langle ... \rangle_\alpha$ represents an average over all particles of species $\alpha$ and $d$ is the system dimensionality.
The self diffusivity averaged over all particles (irrespective of species), $D^{\rm self}$ is found using Eq.~\eqref{msddiffusion} (see Fig.~\ref{sfig:msd}) with the average now taken over all particles.
All self diffusion coefficients presented in the main text are obtained from the MSD.

Self diffusivity can alternatively be computed from a Green-Kubo (G-K) relation: 
\begin{equation}
    \label{gkdiffusiongeneral}
    D^{\rm self} = \frac{1}{d}\int_{t_0}^{\infty}dt \left\langle\mathbf{\dot{x}}(t+t_0)\cdot\mathbf{\dot{x}}(t_0)\right\rangle
\end{equation}
where the average is taken over all particles and initial times. 
Equation~\eqref{gkdiffusiongeneral} is typically derived beginning from the MSD and invoking time-reversal symmetry.
Intrinsically nonequilibrium systems, such as the system explored here, violate time-reversal symmetry. 
However, the G-K relation can be derived more generally~\cite{Hargus2021} without invoking time-reversal symmetry.

The advantage of Eq.~\eqref{gkdiffusiongeneral} is that, for overdamped dynamics with a constant translational drag coefficient $\zeta$, forces can be directly substituted for velocities as the velocity of particle $i$ is simply $\mathbf{\dot{x}} = \mathbf{F}_i/\zeta$. 
As a result, we can express the G-K relation as:
\begin{equation}
    \label{gkdiffusion}
    D^{\rm self} = \frac{1}{d\zeta^2}\int_0^{\infty}dt C(t)
\end{equation}
where $C(t) = \langle \mathbf{F}(t+t_0)\cdot \mathbf{F}(t_0)\rangle$ is a force-force time-correlation function, where $\langle...\rangle$ represents an average over all particles, and the system is assumed to be isotropic~\cite{Hargus2021}.
$C(0)$ represents the stationary variance of the force correlations. 
We are now positioned to understand the precise contributions of various forces to particle mobility.

\begin{figure}
	\centering
	\includegraphics[width=.7\textwidth]{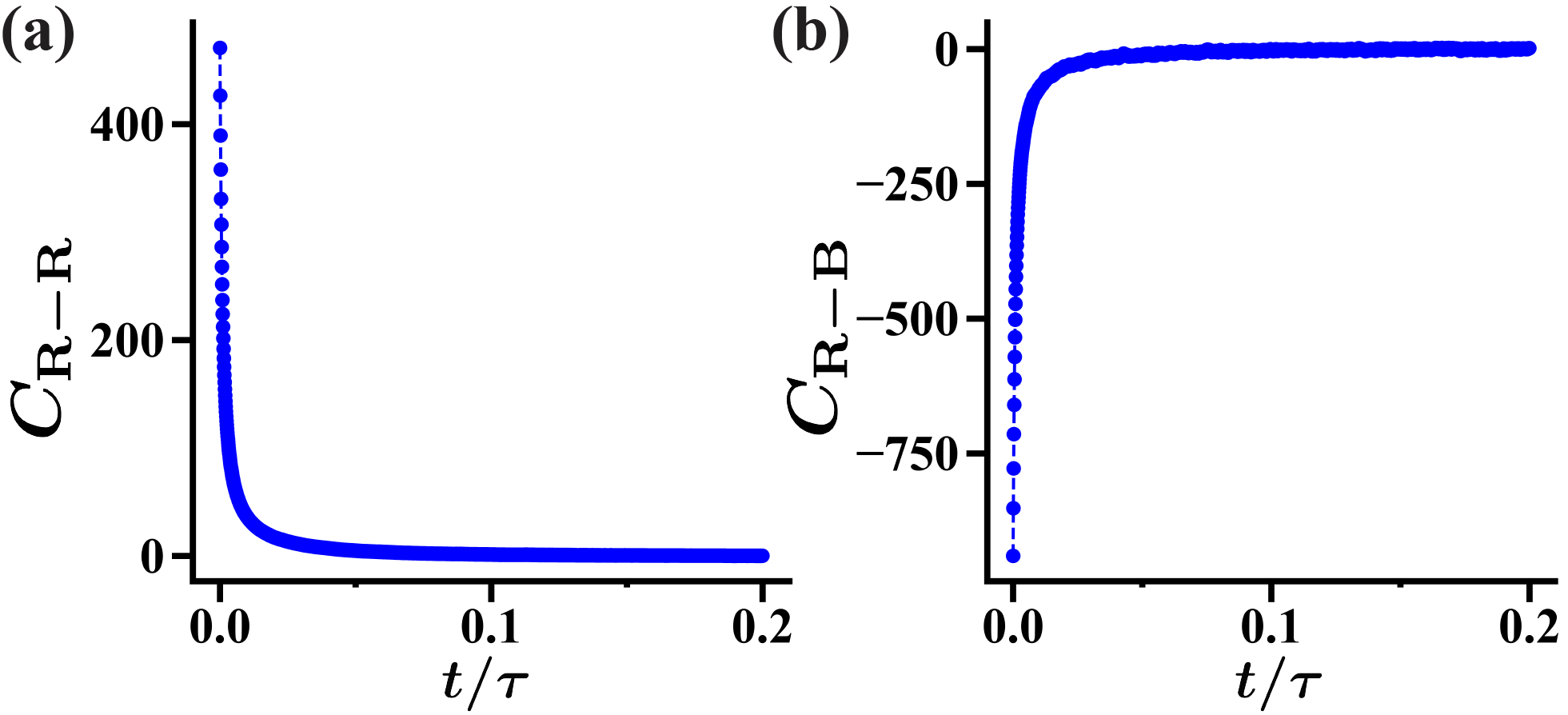}
	\caption{\protect\small{{Components of force correlation (a) $C_{\rm R-R}$ and (b) $C_{\rm R-B}$ in an equilibrium WCA system with $\phi = 0.4$.}}}
	\label{sfig:eqforcecorrelations}
\end{figure}

It is instructive to first consider passive systems before proceeding to our driven system. 
In the reciprocal limit ($\Delta = 0$), there are only two forces acting on particles: the translational Brownian force and reciprocal interparticle force, $\mathbf{F} = \mathbf{F}^{\rm R}+\mathbf{F}^{\rm B}$.
The reciprocal interaction force for particle $i$ is simply $\mathbf{F}_i^{\rm R} = \sum_{j \neq i} \mathbf{F}^{\rm R}_{ij}$.
The force-correlation function $C(t)$ thus has four additive contributions: $C_{R-R}(t) = \langle\mathbf{F}^R(t)\cdot\mathbf{F}^R(0)\rangle$, $C_{B-B}(t) = \langle\mathbf{F}^B(t)\cdot\mathbf{F}^B(0)\rangle$, $C_{R-B}(t) =\langle \mathbf{F}^R(t)\cdot\mathbf{F}^B(0)\rangle$, and $C_{B-R}(t) =\langle \mathbf{F}^B(t)\cdot\mathbf{F}^R(0)\rangle$. 
$C_{B-B}(t) = 2dD_T\zeta^2\delta(t)$ is found analytically from the statistics of the stochastic force and $C_{B-R}(t) = 0$ since the stochastic force, by definition, has no correlation with past forces.
Therefore, there are two non-trivial correlations for passive systems: $C_{\rm R-R}$ and $C_{\rm R-B}$.
We compute these correlations for a collection of repulsive (using the Weeks-Chandler-Anderson (WCA) potential~\cite{Weeks1971} with $\epsilon/k_BT = 2$) particles at a volume fraction of $\phi = \rho\pi(2^{1/6}\sigma)^3/6 = 0.4$.
At these conditions, the system is a homogeneous equilibrium fluid. 
The correlation functions are presented in Fig.~\ref{sfig:eqforcecorrelations}. 
$C_{\rm R-R}$  is strictly positive: reciprocal force correlations increase the particle mobility. 
However, the future reciprocal force and past stochastic force are strongly anticorrelated (i.e., $C_{\rm R-B} < 0$) and outweigh the positive contribution of $C_{\rm R-R}$.  
The anticorrelation between reciprocal forces and past stochastic forces can be understood through a simple picture: in a crowded, high $\phi$ scenario, when $\mathbf{F}^B$ pushes a particle in a particular direction, creating a local density asymmetry around the particle. 
Conservative forces then act to push the particle back to its original location, with a direction opposite to that of the original stochastic force. 
The $D^{\rm self}$ for this reciprocal system is thus, consistent with our expectations, strictly less than the ideal translational diffusivity $D_T$ that is generated by $C_{B-B}$.
The self diffusion (computed both using the G-K relation and MSD with exact agreement) is found to be $D^{\rm self}/D_T = 0.418$.

\begin{figure*}
	\centering
	\includegraphics[width=.95\textwidth]{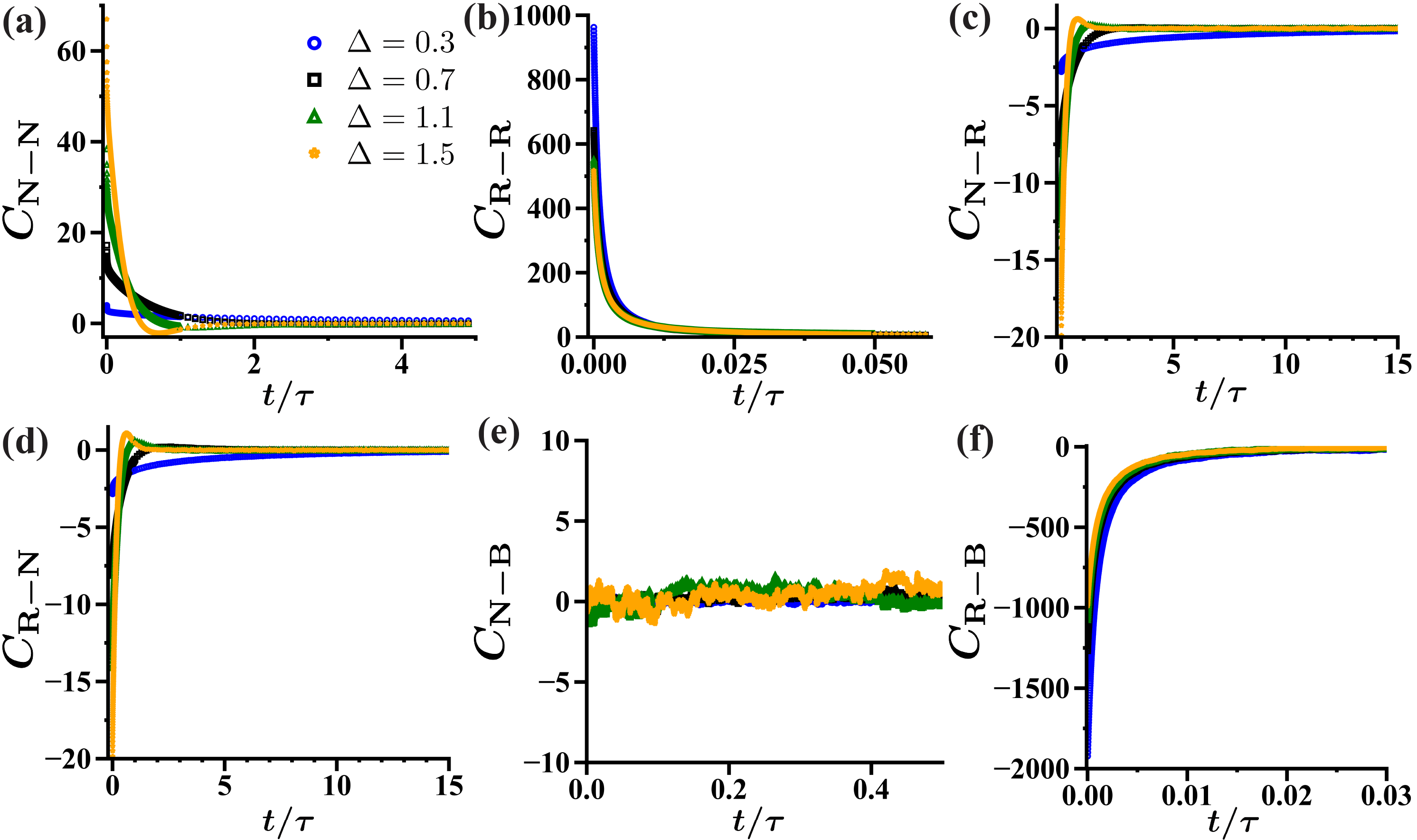}
	\caption{\protect\small{{Nonreciprocity dependence of the force time correlations in the homogeneous clustering regime ($\phi = 0.2$) with: (a) $C_{\rm N-N}$, (b) $C_{\rm R-R}$, (c) $C_{\rm N-R}$, (d) $C_{\rm R-N}$, (e) $C_{\rm N-B}$ and (f) $C_{\rm R-B}$}.}}
	\label{sfig:forcecorrelations}
\end{figure*}

We now compute the correlation functions for our nonreciprocal system, which now have the additional nonreciprocal force such that the (non-drag) forces acting on each particle are, $\mathbf{F} = \mathbf{F}^{\rm R}+\mathbf{F}^{\rm NR}+\mathbf{F}^{\rm B}$ (where the nonreciprocal force on particle $i$ is $\mathbf{F}_i^{\rm NR} = \sum_{j \neq i} \mathbf{F}^{\rm NR}_{ij}$).  
The nonreciprocal $C(t)$ can be decomposed into the nine additive contributions: $C_{\rm N-N}(t) = \langle \mathbf{F}^{\rm NR}(t)\cdot\mathbf{F}^{\rm NR}(0)\rangle$, $C_{\rm R-R}(t) = \langle \mathbf{F}^{\rm R}(t)\cdot\mathbf{F}^{\rm R}(0)\rangle$, $C_{\rm N-R}(t) = \langle \mathbf{F}^{\rm NR}(t)\cdot\mathbf{F}^{\rm R}(0)\rangle$, $C_{\rm R-N}(t) = \langle \mathbf{F}^{\rm R}(t)\cdot\mathbf{F}^{\rm NR}(0)\rangle$, $C_{\rm N-B}(t) = \langle \mathbf{F}^{\rm NR}(t)\cdot\mathbf{F}^{\rm B}(0)\rangle$, $C_{\rm R-B}(t) = \langle \mathbf{F}^{\rm R}(t)\cdot\mathbf{F}^{\rm B}(0)\rangle$, $C_{\rm B-N}(t) = \langle \mathbf{F}^{\rm B}(t)\cdot\mathbf{F}^{\rm NR}(0)\rangle$, $C_{\rm B-R}(t) = \langle \mathbf{F}^{\rm B}(t)\cdot\mathbf{F}^{\rm R}(0)\rangle$, $C_{\rm B-B}(t) = \langle \mathbf{F}^{\rm B}(t)\cdot\mathbf{F}^{\rm B}(0)\rangle$. 
Just as in the case of the passive system, the stochastic Brownian force has no correlation with past forces and thus $C_{\rm B-N}(t) = C_{\rm B-R}(t) = 0$. 
Additionally, once again $C_{B-B}(t) = 2dD_T\zeta^2\delta(t)$.
There are thus six force-force correlation functions that need to be computed for nonreciprocal systems.
We again focus a global density of $\phi = 0.2$ and compute these six correlation functions as a function of $\Delta$ with the results shown in Fig.~\ref{sfig:forcecorrelations}.

Physically, it is initially tempting to attribute the $\Delta$ dependence of the enhanced self diffusivity to be directly driven by the nonreciprocal forces which directly scale with $\Delta$.
The direct contribution of nonreciprocal forcing is thus ${D_{\rm N} = \frac{1}{d\zeta^2}\int_0^{\infty} dt C_{\rm N-N}(t)}$.
The $\Delta$ trend of the direct contribution of nonreciprocal forcing to self diffusivity, $D_{\rm N}$, is presented in Fig.~\ref{sfig:diffusivities}(a).
While we find that $D_{\rm N}$ is always greater than the ideal diffusion $\left(D_{\rm N}/D_T \gtrsim 2\right)$, intriguingly, it is a slightly decreasing function of $\Delta$. 
This can be understood by separately considering the structural and dynamical contributions to $D_{\rm N}$. 

\begin{figure}
	\centering
	\includegraphics[width=.95\textwidth]{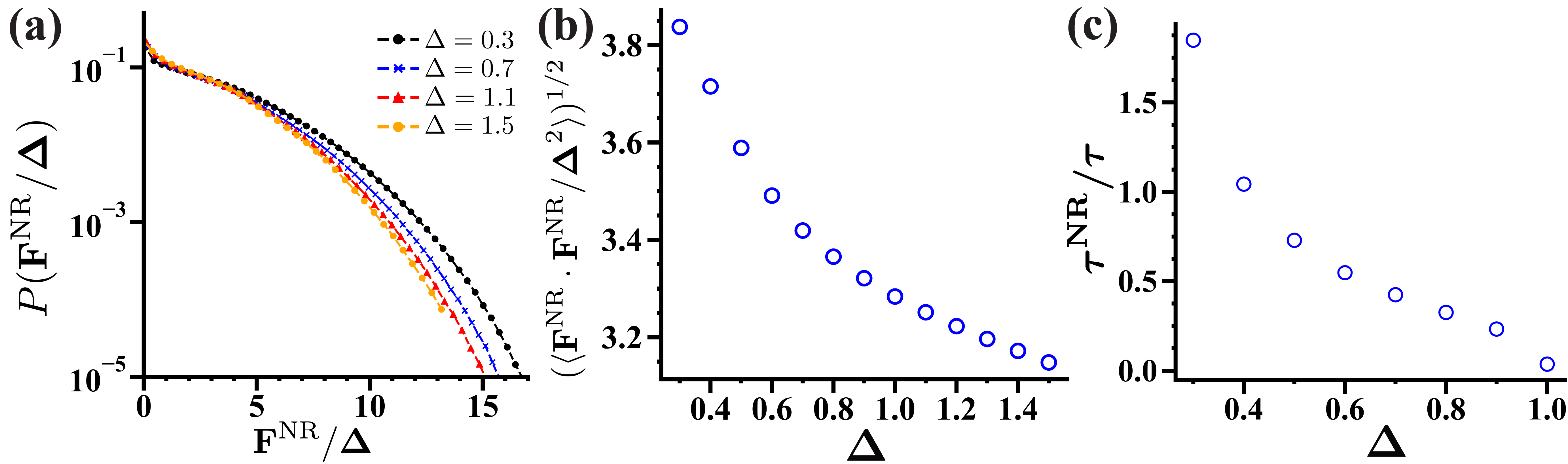}
	\caption{\protect\small{{Nonreciprocity dependence of the (a) probability of $\mathbf{F}^{\rm NR}/\Delta$ (symmetric about $\mathbf{F}^{\rm NR} = \mathbf{0}$); (b) variance of the nonreciprocal force obtained from (a); and (c) nonreciprocal force correlation ($C_{\rm N-N}$) decay time. For all data, the global density is $\phi - 0.2$.}}}
	\label{sfig:nrdiffusion}
\end{figure}

Figure~\ref{sfig:nrdiffusion} provides the distribution of $\mathbf{F}^{\rm NR}$ as a function of $\Delta$. 
For lower values of $\Delta$, the temporal correlations of the nonreciprocal force time correlations are found to be well-fit by an exponential decay with a decay time of $\tau^{\rm NR}$ [see Fig.~\ref{sfig:nrdiffusion}(c)], resulting in ${D_{\rm N} = \langle \mathbf{{F}^{\rm NR}}\cdot \mathbf{{F}^{\rm NR}}\rangle\tau^{\rm NR}/(3\zeta^2)}$ where $\langle \mathbf{{F}^{\rm NR}}\cdot \mathbf{{F}^{\rm NR}}\rangle$ is the stationary variance of the nonreciprocal force felt by a particle [see Fig.~\ref{sfig:nrdiffusion}(a), (b)]. 
This variance is related to the interparticle spacing and the composition asymmetry around a tagged particle [see Fig.~3(c) in main text] along with the strength of the nonreciprocal force. 
If the distribution of microscopic particle configurations is relatively unaltered with increasing $\Delta$, then we anticipate the variance to scale as $\langle \mathbf{{F}^{\rm NR}}\cdot \mathbf{{F}^{\rm NR}}\rangle \sim \Delta^2$.
In detail, we find $\langle \mathbf{{F}^{\rm NR}}\cdot \mathbf{{F}^{\rm NR}}\rangle \sim \Delta^{1.75}$, indicating that these microscopic configurations are altered in a manner that slightly reduces nonreciprocal forcing with increasing $\Delta$.
However, the timescale for rearrangements, $\tau^{\rm NR}$, is found to rapidly decay with increasing nonreciprocity, with  $\tau^{\rm NR} \sim \Delta^{-2.5}$ at lower $\Delta$ and $\Delta^{-1.37}$ at higher $\Delta$. 
As result, $D_{\rm N}$ is only weakly dependent on $\Delta$ and is not driving the observed enhanced diffusion. 

\begin{figure}
	\centering
	\includegraphics[width=.7\textwidth]{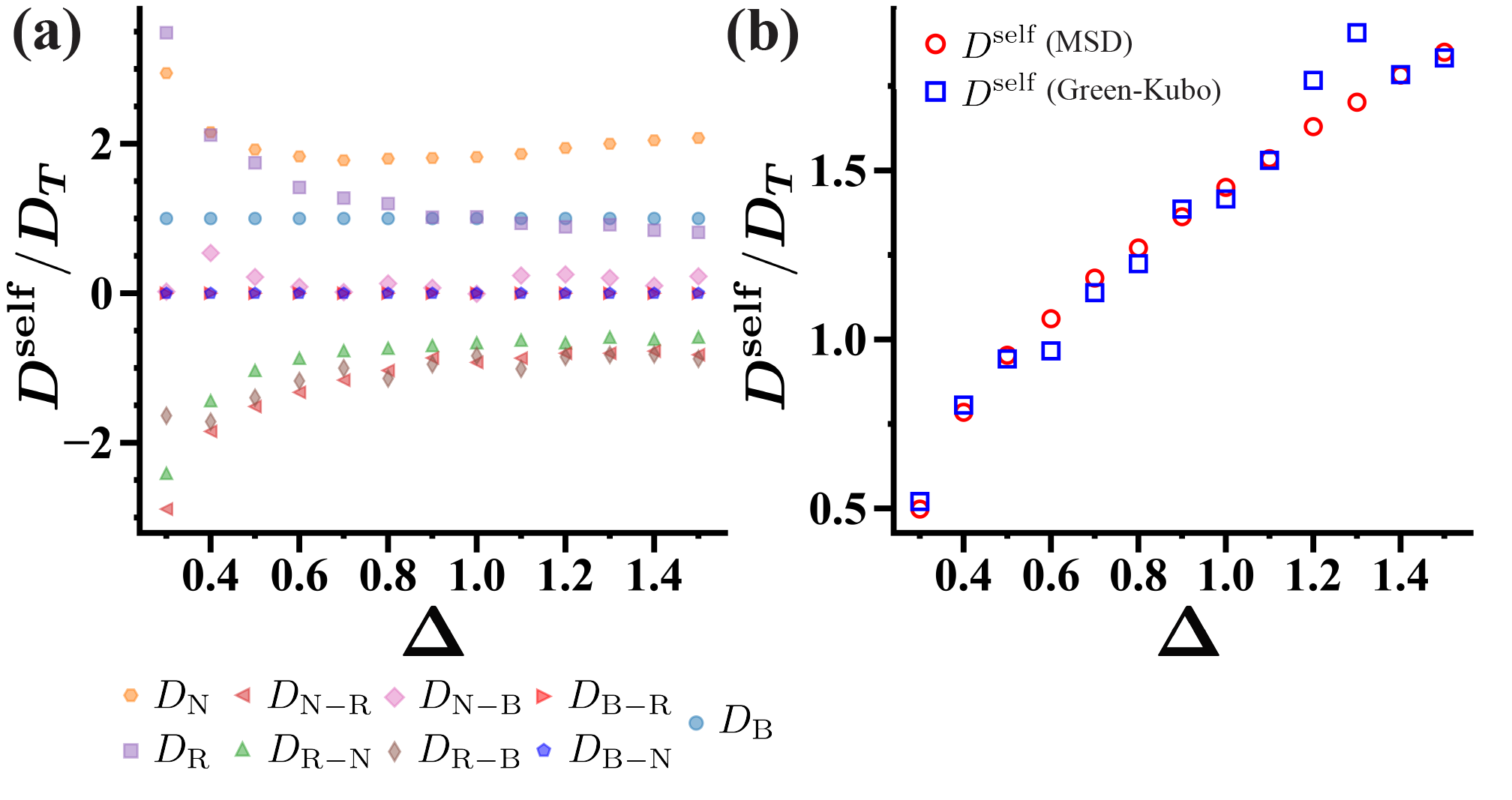}
	\caption{\protect\small{{(a) Components of self diffusion coefficient obtained by integrating the various force correlation functions. (b) Comparison of $D^{\rm self}$ obtained from the MSD and G-K relation. }}}
	\label{sfig:diffusivities}
\end{figure}

What then is driving the enhanced diffusion with increasing $\Delta$? 
To determine this, we now examine the other force correlations. 
From Fig.~\ref{sfig:forcecorrelations}, we can immediately note that $C_{\rm N-N}$, $C_{\rm N-R}$ and $C_{\rm R-N}$ decay over much longer timescales in comparison to $C_{\rm R-R}$ and $C_{\rm R-B}$, which have comparable and fast decay times.
That correlations involving $\mathbf{F}^{\rm NR}$ necessarily appear to decay over longer times, for all $\Delta$ examined here, is intriguing. 
We also observed a negative correlation between reciprocal and nonreciprocal forces, with $C_{\rm N-R}<0$ [Fig.~\ref{sfig:forcecorrelations}(c)] and $C_{\rm R-N}<0$ [Fig.~\ref{sfig:forcecorrelations}(d)]. 
As the reciprocal and nonreciprocal forces are in opposing directions for particles of species $L$ (and the same direction for particles of species $G$), this suggests that a combination of the composition asymmetry and interparticle distances around a tagged $L$ particle differs from that of a tagged $G$ particle.

These negative correlations, which depend intimately on the local structure, act to localize a particle and reduce its self diffusivity. 
In this sense, and despite the presence of nonreciprocal forces, they are similar to the ``frictional'' force correlations that reduce passive particle motion. 
The contribution of each force correlation to the self diffusion is presented in Fig.~\ref{sfig:diffusivities}(a). 
With increasing $\Delta$, the diminished magnitude of the contributions from nonreciprocal-reciprocal force correlations and the correlations between the Brownian and reciprocal force drive the observed $\Delta$ dependence of diffusivity. 
The homogenization of the structure (see Fig.~\ref{sfig:rdf}) reduces the magnitude of these force correlations as the local structure around $L$ and $G$ particles become more similar (eliminating the nonreciprocal-reciprocal force correlations) and less dense (reducing the Brownian and reciprocal force correlations). 
Finally, we show the self diffusion obtained from our force autocorrelations and the Green-Kubo relation [Eq.~\eqref{gkdiffusiongeneral}] agrees with that obtained $D^{\rm self}$ [Eq.~\eqref{msddiffusion}] in Fig.~\ref{sfig:diffusivities}(b).